\documentclass[journal,twocolumn]{IEEEtran}
\usepackage{amsmath,amsfonts}
\usepackage{algorithmic}
\usepackage{algorithm}
\usepackage{array}
\usepackage[caption=false,font=normalsize,labelfont=sf,textfont=sf]{subfig}
\usepackage{textcomp}
\usepackage{stfloats}
\usepackage{url}
\usepackage{verbatim}
\usepackage{graphicx}
\usepackage{cite}
\usepackage{amssymb}
\usepackage{bm}
\usepackage{bbm}
\usepackage{mathrsfs}  
\usepackage{diagbox}
\usepackage{multirow}
\usepackage{graphicx}
\usepackage{color}
\usepackage{tablefootnote}

\newtheorem{theorem}{Theorem}
\newtheorem{lemma}{Lemma}
\newtheorem{remark}{Remark}

\definecolor{darkred}{RGB}{150,50,50}
\definecolor{brown}{RGB}{250,100,100}
\definecolor{green}{RGB}{000,150,100}
\definecolor{navy}{RGB}{000,000,100}

\def\RR{\mathbb{R}}

\def \diff {\mathrm{diff}}

\def \Pp {{\mathbbm P}}
\def \Ep {{\mathbbm E}}
\def \Vp {{\rm V a r }}
\def \Covp {{\rm C o v}}
\def \Cov {\mathbf{ C o v}}

\def \Pb {{\mathbbm P}}

\def\indi{\mathbbm{1}}

\def \A {\mathbf{A}}
\def \B {\mathbf{B}}

\def \I {\mathbf{I}}

\def \V {\mathbf{V}}

\def \R {\mathbf{R}}

\def \bSigma {\mathbf{\Sigma}}

\def \Mg{\rm M_g}

\def\cG{\mathcal{G}}

\def\trans{^{\scriptscriptstyle \sf T}}

\def\bzero{\mathbf{0}}
\def \bone{\mathbf{1}}
\def \bmu{\boldsymbol{\mu}}
\def \bSigma{\boldsymbol{\Sigma}}

\begin{document}

\title{Asymptotic Distribution-free Change-point Detection for Modern Data Based on a New Ranking Scheme}

\author{Doudou Zhou and Hao Chen
        % <-this % stops a space
\thanks{This work of Doudou Zhou was supported in part by the NSF award DMS-1848579 and NUS Start-up Grant A-0009985-00-00. The work of Hao Chen was supported in part by the NSF awards DMS-1848579 and DMS-2311399. This paper was presented in part at the 2024 Joint Statistical Meetings (JSM) and in part at the 18th International Joint Conference on Computational and Financial Econometrics and Computational and Methodological Statistics (CFE-CMStatistics 2024). \emph{(Corresponding author: Hao Chen.)} \\
\indent Doudou Zhou is now with the Department of Statistics and Data Science, National University of Singapore, 117546 Singapore (e-mail: ddzhou@nus.edu.sg). \\
\indent Hao Chen is now with the Department of Statistics, University or California, Davis, CA 95616 USA (e-mail: hxchen@ucdavis.edu).}        
%\thanks{This paper was produced by the IEEE Publication Technology Group. They are in Piscataway, NJ.}% <-this % stops a space
%\thanks{Manuscript received April 19, 2021; revised August 16, 2021.}
}

% The paper headers
\markboth{Journal of \LaTeX\ Class Files,~Vol.~14, No.~8, August~2021}%
{Shell \MakeLowercase{\textit{et al.}}: A Sample Article Using IEEEtran.cls for IEEE Journals}

%\IEEEpubid{0000--0000/00\$00.00~\copyright~2021 IEEE}
% Remember, if you use this you must call \IEEEpubidadjcol in the second
% column for its text to clear the IEEEpubid mark.

\maketitle

\begin{abstract}
Change-point detection (CPD) involves identifying distributional changes in a sequence of independent observations. Among nonparametric methods, rank-based methods are attractive due to their robustness and effectiveness, and have been extensively studied for univariate data. However, they are not well explored for high-dimensional or non-Euclidean data. This paper proposes a new method, Rank INduced by Graph Change-Point Detection (RING-CPD), which utilizes graph-induced ranks to handle high-dimensional and non-Euclidean data. The new method is asymptotically distribution-free under the null hypothesis, and an analytic $p$-value approximation is provided for easy type-I error control. Simulation studies show that RING-CPD effectively detects change points across a wide range of alternatives and is also robust to heavy-tailed distribution and outliers. The new method is illustrated by the detection of seizures in a functional connectivity network dataset, changes in digit images, and travel pattern changes in the New York City Taxi dataset.
\end{abstract}


\section{Introduction}

Given a sequence of independent observations, an important problem is to decide whether the observations are from the same distribution or there is a change of distribution at a certain time point. Change-point detection (CPD) has attracted a lot of interest since the seminal work of \cite{page1954continuous}, { evolving into two primary strands: online (or sequential) CPD, which aims to detect distributional changes in real-time as data flows in (see for example \cite{desobry2005online, tartakovsky2008asymptotically, xie2013sequential, chen2019sequential,xie2020sequential,chu2022sequential}), and offline CPD, which involves analyzing a completely observed data sequence  and is the focal point of this paper.}  In this big data era, CPD has diverse applications in many fields, including functional magnetic resonance recordings \cite{barnett2016change,zambon2019change}, healthcare \cite{staudacher2005new,malladi2013online}, communication network evolution \cite{kossinets2006empirical,eagle2009inferring, peel2015detecting}, and financial modeling \cite{bai1998estimating,talih2005structural}. Parametric approaches (see for example, \cite{srivastava1986likelihood, zhang2010detecting,siegmund2011detecting,chen2012parametric,wang2018change}) are useful to address the problem for univariate and low-dimensional data, however, they are limited for high-dimensional or non-Euclidean data due to a large number of parameters to be estimated unless strong assumptions are imposed.

A few nonparametric methods have been proposed, including kernel-based methods  \cite{desobry2005online, li2015m, garreau2018consistent, arlot2019kernel,chang2019kernel},  interpoint distance-based methods \cite{matteson2014nonparametric,li2020asymptotic} and graph-based methods \cite{chen2015graph, shi2017consistent,chu2019asymptotic,chen2019change,song2020asymptotic,liu2020fast,nie2021weighted, chen2023graph, chu2025tightness}. When dealing with high-dimensional data, the curse of dimensionality can pose a huge problem. The kernel-based method \cite{li2015m}  assumes that the high-dimensional data resides on a low-dimensional manifold to get around this problem. On the other hand, the graph-based methods  \cite{chu2019asymptotic} incorporate a useful pattern caused by the curse of dimensionality \cite{chen2017new}, and are effective in analyzing high-dimensional and non-Euclidean data for a wide range of changes. However, the graph-based methods have focused on unweighted graphs, which may cause information loss. Later, \cite{li2020asymptotic} adopted a similar idea and proposed an asymptotic distribution-free approach utilizing all interpoint distances that worked well for detecting both location and scale changes. However, their test statistics are time and memory-consuming and implicitly require the existence of the second moment of the underlying distribution, which can be violated by heavy-tailed data or outliers that are common in many applications.

Among the nonparametric methods, rank-based methods are attractive for univariate data due to their robustness and effectiveness 
\cite{bhattacharyya1968nonparametric, darkhovskh1976nonparametric, pettitt1979non, schechtman1982nonparametric, lombard1987rank,lombard1983asymptotic, gerstenberger2018robust, wang2020rank}. However, they are less explored for high-dimensional or non-Euclidean data. Specifically, existing multivariate rank-based methods are limited in many ways. For instance, \cite{lung2015homogeneity} proposed to use the component-wise rank, which requires the dimension of the data to be smaller than the number of observations and suffers from dependent covariates. \cite{zhang2020spatial} and \cite{shu2022spatial} proposed the spatial rank-based methods, which were designed mainly for detecting mean shifts. \cite{chenouri2020robust} proposed to use the ranks obtained from data depths, which is often used for low-dimensional data and is computationally intensive when the dimension is high.

Noticing the gap between the potential benefit of the rank-based method and the scarce exploration for multivariate/high-dimensional data, we propose a new rank-based method called {\bf R}ank {\bf IN}duced by {\bf G}raph {\bf C}hange-{\bf P}oint {\bf D}etection (RING-CPD), which can be applied to high-dimensional and non-Euclidean data. Unlike previous works dealing with the ranks of observations that are often limited to low-dimensional distributions,  we propose to use the rank induced by similarity graphs that can be applied to data whose dimension could be much larger than the sample size.
The graph-induced rank \cite{rise} is the rank defined in the similarity graphs. Instead of treating all edges in the graph equally, we assign the rank as weights to each edge and construct the scan statistic based on the ranks. Discussions on this rank and the new test are presented in Section \ref{sec:method}.  We prove that the proposed scan statistic is asymptotically distribution-free, facilitating its usage to a broader community of researchers and analysts. They are also consistent against all types of changes under certain conditions of the similarity graph (Section \ref{sec: asymptotic}).  The proposed statistic can work for a wide range of alternatives and is robust to heavy-tailed distributions and outliers, as illustrated by extensive simulation studies in Section \ref{sec: simulation} and three real data examples in Section \ref{sec: real data}.  The details of proofs of theorems are deferred to the Supplementary Material.

\section{Method}
\label{sec:method}
For a sequence of independent $\{y_{i} \}_{i=1}^n$, we consider testing 
$$
H_{0}: y_i \sim F_{0}, \quad i=1, \ldots, n 
$$
against the single change-point alternative
\begin{equation}
 H_{1}: \exists 1 \leq \tau<n, y_{i} \sim \begin{cases} F_{0}, & i \leq \tau, \\ F_{1}, & \text { otherwise }\end{cases}
 \label{H1}
\end{equation}
or the changed interval alternative
\begin{equation}
    H_{2}: \exists 1 \leq \tau_{1}<\tau_{2} \leq n,  y_{i} \sim \begin{cases} F_{1}, & i\in[\tau_{1}+1, \tau_{2}] \\ F_{0}, & \text {otherwise,}\end{cases}
\label{H2}
\end{equation}
where $F_{0}$ and $F_{1}$ are two different distributions. When there are multiple change points, the two alternatives can be applied recursively. Alternatively, our method in the following can be extended similarly to \cite{zhang2021graph} to accommodate multiple change-points, using the idea of wild binary segmentation \cite{fryzlewicz2014wild} or seeded binary segmentation \cite{kovacs2020seeded}.

\begin{figure*}
 \centering
 \begin{tabular}{cc}          \includegraphics[width=0.35\textwidth]{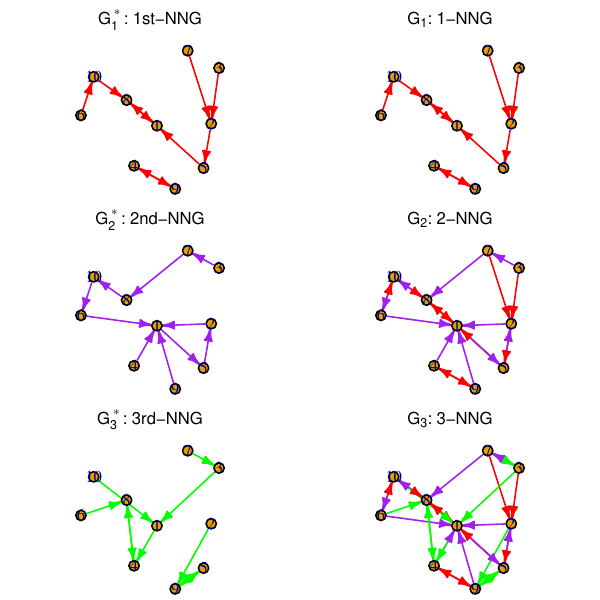} & 
           \includegraphics[width=0.35\textwidth]{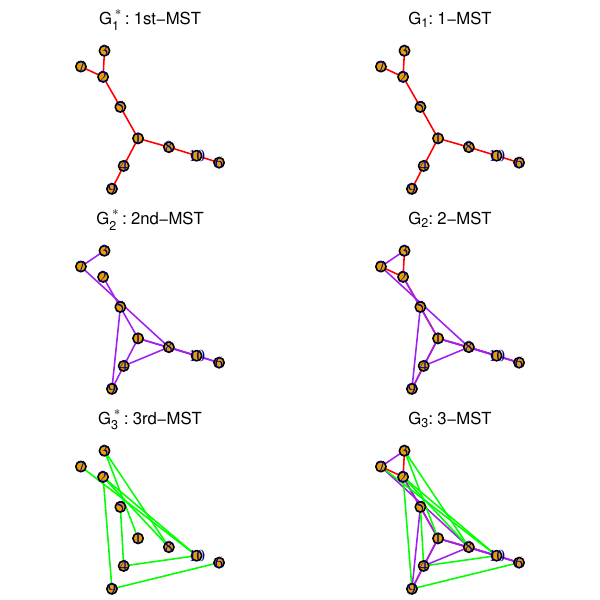} \\    
(i) $k$-NNG. & (ii) $k$-MST.  \\
\end{tabular}
     \caption{Examples of different similarity graphs.}
     \label{fig:graph}
 \end{figure*}

 Our methodology begins by transforming the observations $\{y_{i}\}_{i=1}^n$ into a graph-induced rank matrix $\mathbf{R} = (R_{ij})_{i,j=1}^n \in \RR^{n \times n}$, which encapsulates the similarity between pairs of observations. Then we derive scan statistics utilizing $\mathbf{R}$. We follow notations in \cite{rise}.  Let $V = \{1,2,\ldots,n\}$ denote the vertex set of the $n$ observations. A graph $G$ on these observations is defined as $G = (V,E)$, where $E \subseteq V \times V$. We define that an edge $(i,j) \in G$ if $(i,j) \in E$. 
 For two graphs, $G_1 = (V,E_1)$ and $G_2 = (V,E_2)$, sharing the same vertex set $V$, we define their intersection as  $G_1 \cap G_2 = (V,E_1 \cap E_2)$, which is empty when there are no shared edges, i.e., $G_1 \cap G_2 = \emptyset$ if $E_1 \cap E_2 =\emptyset$, and their union as $G_1 \cup G_2 = (V,E_1 \cup E_2)$. Given $\{y_{i}\}_{i=1}^n$, we sequentially construct a series of simple similarity graphs\footnote{A simple graph is a graph without self-loops and multiple edges between any two 
vertices.}  $\{ G_l \}_{l=0}^k$, starting with an edgeless graph  $G_0 = (V, \emptyset)$. 
The process continues with
$G_{l+1} = G_{l} \cup G_{l+1}^{*},$ 
where $G_{l+1}^{*} = \arg \max_{G' \in \cG_{l+1} } \sum_{(i,j) \in G'} S( y_i,y_j) \text{ and } \cG_{l+1} = \{G' \in \cG: G'\cap G_{l} =  \emptyset\}.$
Here, $\mathcal{G}$ represents a set of graphs subject to certain structural constraints, and $S(\cdot,\cdot)$ is a similarity measure, such as $S(y_i,y_j) = -\|y_i - y_j\|$ for Euclidean data, with $\|\cdot\|$ denoting the Euclidean norm. This framework allows for the construction of various well-established similarity graphs under different constraints. For instance, both the $k$-nearest neighbor graph ($k$-NNG) and the $k$-minimum spanning tree ($k$-MST)\footnote{The MST is a spanning tree that minimizes the sum of distances of edges in the tree while connecting all observations. The $k$-MST is the union of the $1$st, \ldots, $k$th MSTs, where the $k$th MST is a spanning tree that connects all observations while minimizing the sum of distances across edges excluding edges in the $(k-1)$-MST.}\cite{friedman1979multivariate} satisfy the above definition of the sequence of graphs. For NNG, the structural constraint on $G^\prime$ is that each vertex $i$ connects to exactly one other vertex.  Specifically, the graph set $\mathcal{G}$ is defined as $\mathcal{G} = \{ G' = (V,E') :$ for each $i$, there exists one and only one $j \in V\backslash \{i\}$ such that $(i,j) \in E'\}$. Thus, $G_1$ corresponds to the $1$-NNG, where each vertex $i$ connects to the vertex $j$ with the largest similarity $S(y_i, y_j)$ among all other observations. Similarly, $G_{l}$ is the $l$-NNG,  $G_{l+1}^*$ is the $(l+1)$th NNG and $G_{l+1}$ is the $(l+1)$-NNG.  For $k$-MST, the constraint is that $G^\prime$ must be a tree connecting all vertices, making $G_{l}$ the $l$-MST, $G_{l+1}^*$ the $(l+1)$th MST and $G_{l+1}$ the $(l+1)$-MST.  An illustration of these graphs is presented in Figure~\ref{fig:graph}. For more choices of graphs, one can see \cite{rise}.

\begin{figure*}[!t]
\begin{center}
\begin{tabular}{c}
     \includegraphics[width=0.75\textwidth]{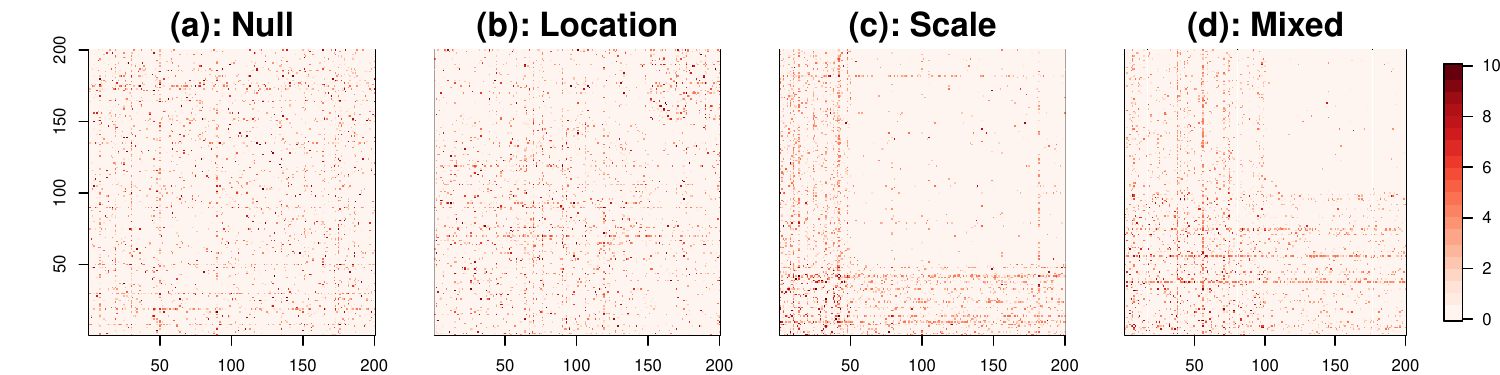}  \\
 \includegraphics[width=0.75\textwidth]{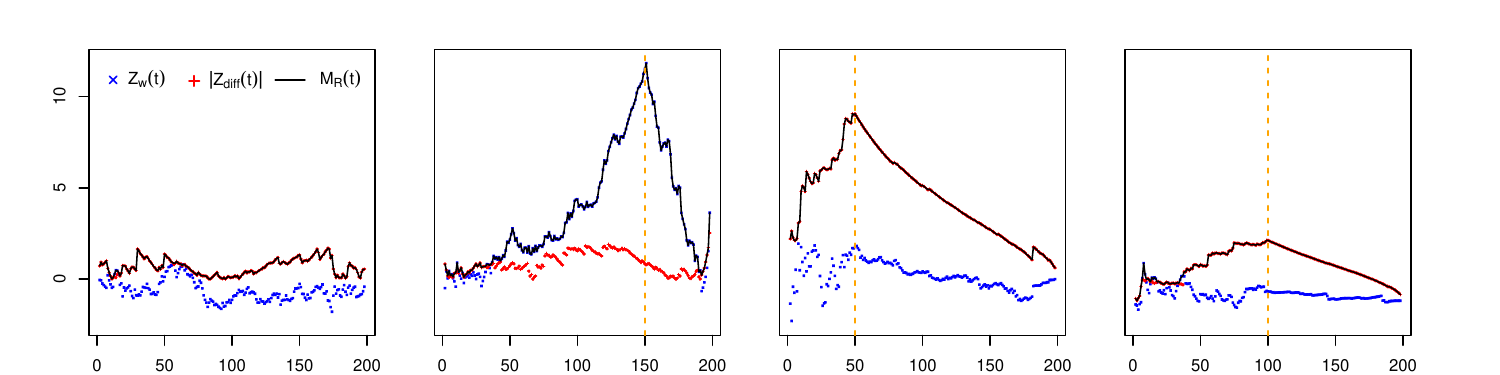} 
\end{tabular}
\end{center}
\caption{Top panel: heatmap of the graph-induced rank matrix $\R$ in $10$-NNG. Bottom panel: the values of $Z_w(t) = Z_{w}(0, t)$ (blue), $|Z_{\diff}(t)| = |Z_{\diff}(0, t)|$ (red) and $M_R(t)$ (black) against $t$ under different scenarios. The true change points are denoted by the vertical dashed lines (orange). }
\label{fig:rank}
\end{figure*}

With $\{ G_l \}_{l=1}^k$, the graph-induced rank matrix  is defined as 
\[ R_{i j } = \sum_{l=1}^k \indi\big( (i,j) \in G_l \big) \text{ for } 1 \leq i,j \leq n   \,,\]
where for an event $A$, $\indi(A)$ is an indicator function that equals one if event $A$ occurs, and equals zero otherwise. The graph-induced ranks impose more weights on the edges with higher similarity, thus incorporating more similarity information than the unweighted graph. In the meantime, the robustness property of the ranks makes the weights less sensitive to outliers compared to directly using distance. The behavior of the graph-induced rank matrix in the $10$-NNG is illustrated in Figure~\ref{fig:rank} by simulated data. { The graph-induced rank depends implicitly on $k$, and we will discuss its choice in Section \ref{sec:choice}.}

\begin{remark}[Computational efficiency]
The computational complexities for constructing $k$-NNG and $k$-MST using brute-force methods are $O(n^2d)$ and $O\big(n^2 (d + \log n) \big)$ \cite{friedman1979multivariate}, respectively. More efficient algorithms are widely used in practice. For example, the $k$-NNG can be computed using KD Tree \cite{bentley1975multidimensional} or Ball Tree algorithms \cite{omohundro1989five}, as implemented in scikit-learn \cite{JMLR:v12:pedregosa11a}, with a computational complexity of $O(n \log n)$ when the dimension $d$ is small (e.g., $d \leq 15$ as recommended by scikit-learn package).  However, these methods face challenges due to the ``curse of dimensionality." In such cases, both their empirical and theoretical performance can degrade, and may become comparable or even worse than brute-force algorithms, as discussed in \cite{ram2019revisiting}. On the other hand, the approximate $k$-NNG offers a computational complexity of $O\big( n d(\log n + k \log d) \big)$, even with high-dimensional data \cite{arya1998optimal,ram2019revisiting}.
\end{remark}

\allowdisplaybreaks

 After constructing the graph-induced rank matrix $\mathbf{R}$, we are ready to propose the test statistics. For testing the changed interval alternative $H_2$ \eqref{H2}, each possible interval $(t_1,t_2]$ for $1 \leq t_1 < t_2 \leq n$  partitions the observations into two groups: one group containing all observations observed during $(t_1,t_2]$, and the other group containing all observations observed outside of this
interval. Then, for any candidate changed interval $(t_1,t_2]$, we define two basic quantities: 
\begin{align}
& U_1(t_1,t_2) = \sum_{i=1}^n \sum_{j=1}^n R_{ij} \indi(t_1 < i,j \leq t_2), \\
& U_2(t_1,t_2) = \sum_{i=1}^n \sum_{j=1}^n R_{ij} \indi(i,j \leq t_1 \text{ or } i,j > t_2).      
\end{align}  
We have that $U_1(t_1,t_2)$ is the sum of ranks within the interval $(t_1,t_2]$, and $U_2(t_1,t_2)$ is the sum of ranks outside of the interval $(t_1,t_2]$. Then the max-type two-sample test statistic for testing the changed interval alternative can be defined as 
\begin{equation}
M_R(t_1,t_2) = \max \big(Z_{w}(t_1,t_2), |Z_{\diff}(t_1,t_2)|\big) \,,
\end{equation}
where 
$$
\begin{aligned}
& Z_{w}(t_{1}, t_{2})  =  \frac{U_{w}(t_{1}, t_{2}) - \Ep\big(U_{w}(t_{1}, t_{2})\big)}{\sqrt{\Vp\big(U_{w}(t_{1}, t_{2})\big)}}, \\
& Z_{\diff}(t_{1}, t_{2})  =  \frac{U_{\diff}(t_{1}, t_{2}) - \Ep\big(U_{\diff}(t_{1}, t_{2})\big)}{\sqrt{\Vp\big(U_{\diff}(t_{1}, t_{2})\big)}} \,,    
\end{aligned}
$$
with $ U_{\diff}(t_1,t_2) = U_{1}(t_1,t_2) - U_{2}(t_1,t_2)$ and 
$$U_{w}(t_1,t_2) = \tfrac{n-t_2+t_1-1}{n-2} U_{1}(t_1,t_2) + \tfrac{t_2-t_1-1}{n-2} U_{2}(t_1,t_2).$$
Here we use $\Ep$, $\Vp$, and $\Covp$ to respectively denote the expectation, variance, and covariance under the permutation null distribution, which places $1/n!$ probability on each of the $n!$ permutations of the order of the observations. Under the alternative hypothesis, it is possible that (i) both $U_1(t_1,t_2)$ and $U_2(t_1,t_2)$ are larger than their null expectations (a typical scenario under location alternatives) and (ii)  one of them is larger than while the other one is smaller than its corresponding null expectation (a typical scenario under scale alternatives). See \cite{chen2017new} for more discussions on these scenarios. For (i), $Z_{w}(t_{1}, t_{2})$ will be large and for (ii), $|Z_{\diff}(t_1,t_2)|$ will be large. Thus, $M_R$ is powerful for different types of alternatives.

When testing the single change-point alternative $H_1$ \eqref{H1}, we can simply use $M_R(t) = M_R(0,t)$. To illustrate the behaviors of $Z_{w}$, $Z_{\diff}$ and $M_R$ under different scenarios, we generate $n=200$ independent multivariate observations with dimension $d = 500$ from 
\begin{itemize}
    \item [(a)] (Null) $y_i \sim N(\bzero_d,\bone_d), i =1,\ldots,n$; 
    \item [(b)] (Location shift) $y_i \sim N(\bzero_d,\I_d), i =1,\ldots,3n/4$,  $y_i \sim N(0.21 \bone_d,\I_d), i =3n/4+1,\ldots,n$; 
    \item [(c)] (Scale shift) $y_i \sim N(\bzero_d,\I_d), i =1,\ldots,n/4$,  $y_i \sim N(\bzero_d,1.2 \I_d), i =n/4+1,\ldots,n$; 
    \item [(d)] (Location and scale mixed shift) $y_i \sim N(\bzero_d,\I_d), i =1,\ldots,n/2$,  $y_i \sim N(0.1 \bone_d,1.2 \I_d), i =n/2+1,\ldots,n$.
\end{itemize}
For all numeric experiments in the paper, we use the negative Euclidean norm as the similarity measure unless specifically noted. The values of $Z_w(t) = Z_{w}(0, t)$, $|Z_{\diff}(t)| = |Z_{\diff}(0, t)|$ and $M_R(t)$ against $t$ are presented in Figure~\ref{fig:rank}. 

Although the similarity measure $S$ inherently exhibits symmetry, the rank matrix $\R$ may not always maintain this property, particularly in scenarios where the graph is directed such as the $k$-NNG. For simplicity, we symmetrize $\R$ by $0.5 (\R + \R\trans)$. Note that this symmetrization does not change the value of $U_1$ and $U_2$. Without further specialization, we use $\R$ to denote its symmetrized version in the following. The explicit expressions of $\Ep\big( U_1(t_1,t_2) \big)$, $\Ep\big( U_2(t_1,t_2) \big)$ and  $\bSigma(t_1,t_2)$ are presented in Lemma \ref{lem:moment}. Let 
 $$
 \begin{aligned}
  & \bar  R_{i \cdot} = \frac{ \sum_{j=1}^n R_{ij}}{n-1} \, ,  r_0 = \frac{\sum_{i=1}^n  \bar R_{i\cdot}}{n}  \, , r_1^2 = \frac{\sum_{i=1}^n \bar R_{i \cdot}^2}{n} \, , \\
  &  r_d^2 = \frac{\sum_{i=1}^n \sum_{j=1}^n R_{ij}^2}{n(n-1)}  \, , V_d = r_d^2 - r_0^2 \,, V_r = r_1^2 - r_0^2\,. 
 \end{aligned}
 $$
 \begin{lemma}
\label{lem:moment}
Under the permutation null distribution, we have 
\begin{align*}
&  \Ep\big(U_1(t_1, t_2) \big)   = (t_2 - t_1)(t_2 - t_1 - 1) r_0 \, , \\
&   \Ep\big(U_2(t_1,t_2)\big)  = (n-t_2+t_1)(n-t_2+t_1-1)r_0  \\
&  \Vp \big( U_1(t_1,t_2) \big)   = f_1 (t_2 - t_1) V_d + f_2 (t_2 - t_1)V_r \, , \\
& \Vp \big( U_2(t_1,t_2) \big)  = f_1 (n-t_2+t_1)V_d + f_2 (n-t_2+t_1) V_r \, , \\
 & \Covp \big(U_1(t_1,t_2), U_2(t_1,t_2) \big)   = f_1 (t_2 - t_1) \big(  V_d -  2 (n-1) V_r \big) \,,
\end{align*}
where 
$$
\begin{aligned}
& f_1(t) = \frac{2t(t-1)(n-t) (n-t-1)}{(n-2)(n-3)}, \\ 
& f_2(t) = \frac{4t(n-t)(t-1) (t-2)(n-1)}{(n-2)(n-3)}.    
\end{aligned}
$$
\end{lemma}
The proof of Lemma \ref{lem:moment} is through combinatorial analysis. It can be done similarly to the proof of Theorem $2.1$ in \cite{rise} and thus omitted here. When testing against $H_1$, we reject $H_0$ if the scan statistic
\begin{equation}
\max _{n_{0} \leq t \leq n_{1}} M_R(t)
\end{equation}
exceeds the critical value for a given nominal level.  Here $n_{0}$ and $n_{1}$ are pre-specified integers. A common choice of $n_0$ and $n_1$ is $n_0 = [0.05n]$ and $n_1 = n - n_0$, where $[x]$ denotes the  integer closest to $x$.  { 
When $H_0$ is rejected, the  change-point is  estimated by 
$$\hat \tau = \arg \max _{n_{0} \leq t \leq n_{1}} M_R(t).$$
} 
When testing against $H_2$, we reject $H_0$ if the scan statistic
\begin{equation}
 \max_{1 \leq t_{1}<t_{2} \leq n \atop n_{0} \leq t_{2}-t_{1} \leq n_{1}} M_R(t_1,t_2)    
\end{equation}
exceeds the critical value for a given nominal level. { When $H_0$ is rejected, the  detected changed interval is 
$$(\hat \tau_1, \hat \tau_2) = \arg \max_{1 \leq t_{1}<t_{2} \leq n \atop n_{0} \leq t_{2}-t_{1} \leq n_{1}} M_R(t_1,t_2).$$
}

\begin{remark}
Change-point detection is closely related to two-sample 
hypothesis testing. Our proposed methods utilize two-sample test statistics $M_R$, $Z_w$, $Z_{\rm diff}$ introduced in a prior work \cite{rise}. However, change-point detection is significantly more complicated. For two-sample hypothesis testing, the group labels of the observations are provided. In contrast, for change-point detection, this information is unavailable due to the unknown change-point location. In fact, even the existence of a change-point or not is unknown. Here, we use the maximum of the scan statistic as the test statistic; that is, we compute the two-sample test statistic for each potential change-point location, and we reject the null hypothesis of no change point if any of these statistics exhibit sufficient evidence. Understanding the distribution of the scan statistic is considerably more complicated than its corresponding quantity under the two-sample testing framework. We examine the asymptotic distribution of the scan statistic in Section \ref{sec: asymptotic}. 
\end{remark}

\section{Asymptotic properties of the scan statistics}
\label{sec: asymptotic}

For decision-making, the critical values should be determined. Alternatively, we consider the tail probabilities
\begin{equation}
 \Pp\big( \max _{n_{0} \leq t \leq n_{1}} M_R(t) > b \big)
\label{eq:tail}
\end{equation}
for the single change-point alternative and 
\begin{equation}
 \Pp\big(  \max_{1 \leq t_{1}<t_{2} \leq n \atop l_{0} \leq t_{2}-t_{1} \leq l_{1}} M_R(t_1,t_2)  > b\big)
\end{equation}
for the changed interval alternative, respectively, where $\Pp$ denotes the probability under the permutation null distribution.  When $n$ is small, we can apply the permutation procedure. However, it is time-consuming when $n$ is large. Hence, we derive the asymptotic distribution of the scan statistics for analytic approximations of the tail probabilities.

\subsection{Asymptotic null distributions of the basic processes}

By the definition of $M_R(t)$ and $M_R(t_1,t_2)$, it is sufficient to derive the limiting distributions of 
\begin{equation}
\begin{aligned}
 &   \big\{Z_{\diff}( \lfloor n u \rfloor ): 0<u<1\big\} \text{ and } \\
 & \big\{Z_{w}( \lfloor n u \rfloor ): 0<u<1 \big\}
\end{aligned}
\label{eq:imit1}
\end{equation}
for the single change-point alternative, and 
\begin{equation}
\begin{aligned}
& \big\{ Z_{\diff}(\lfloor n u \rfloor, \lfloor n v \rfloor ): 0<u<v<1 \big\} \text{ and } \\
& \big\{Z_{w}( \lfloor n u \rfloor, \lfloor n v \rfloor ): 0<u<v<1 \big\}      
\end{aligned}
\label{eq:imit2}
\end{equation}
for the changed-interval alternative, where $\lfloor x \rfloor$ denotes the largest integer less than or equal to $x$. We use the notations $a_{n}=o(b_{n})$ and $a_{n} \prec b_{n}$ when $\lim _{n \rightarrow \infty} a_{n}/b_{n} =0$,  and $a_{n} \precsim b_{n}$ when $\lim _{n \rightarrow \infty} a_{n}/b_{n}$ is bounded. The proof of Theorem \ref{thm:basis} is provided in Supplement S1.

\begin{theorem} Under Conditions (1) $r_1 \prec r_d$; (2) $\sum_{i=1}^n\big(\sum_{j=1}^n R_{ij}^2 \big)^2 \precsim n^3 r_d^4$; (3) $\sum_{i=1}^{n} \big|\widetilde R_{i \cdot} \big|^3  \prec ( n V_r)^{1.5}$;  (4) $\sum_{i=1}^n \widetilde R_{i \cdot}^3 \prec n r_d V_r$;    (5) $\big| \sum_{i=1}^n \sum_{j \neq s}^n R_{ij} R_{is}   \widetilde R_{j \cdot} \widetilde R_{s \cdot} \big| \prec n^3 r_d^2 V_r$;  (6) $\sum_{i=1}^n \sum_{j=1}^n  \sum_{
s,l \neq i,j}^n R_{ij} R_{js} R_{sl}R_{li}\prec n^4 r_d^4$, { when $n \rightarrow \infty$}, we have
\begin{enumerate}
    \item $\big\{Z_{\diff}( \lfloor n u \rfloor): 0<u<1\big\}$ and $\big\{Z_{w}( \lfloor n u \rfloor): 0<u<1\big\}$ converge to independent Gaussian processes in finite-dimensional distributions, which we denote as $\big\{Z_{\diff}^{*}(u): 0<u<1\big\}$ and $\big\{Z_{w}^{*}(u): 0<u<1\big\}$, respectively.
    \item  $\big\{Z_{\diff}(\lfloor n u \rfloor, \lfloor n v \rfloor): 0<u<v<1\big\}$ and $\big\{Z_{w}(\lfloor n u \rfloor, \lfloor n v \rfloor): 0<u<v<1\big\}$ converge to independent two-dimension Gaussian random fields in finite dimensional distributions, which we denote as $\big\{Z_{\diff}^{*}(u, v): 0<u<v<1\big\}$ and $\big\{Z_{w}^{*}(u, v): 0<u<v<1\big\}$, respectively.
\end{enumerate}
\label{thm:basis}
\end{theorem}

\begin{remark}
Conditions (1)-(6) put restrictions on the similarity graph and the ranks. Importantly, they require that there should not be an excessive number of hub nodes in the graph and that the variation of the average row-wise ranks, $V_r$, should not be dominated by a small portion of elements. { Specifically, when the largest degree of $G_k$ is bounded by $Ck$ for some constant $C$, Conditions (1), (2), (4), and (6) always hold for $k$-NNG and $k$-MST, and Conditions (3) and (5) also tend to be satisfied as verified by simulation studies. More detailed discussions can be found in Appendix E of \cite{rise}.} 
Note that these conditions are the same as \cite{rise}, which is exciting because we do not need extra conditions when we extend the statistics from the two-sample setting to the change-point detection setting. 
\end{remark}

Let $\rho_{w}^{*}(u, v)= \Cov\big(Z_{w}^{*}(u), Z_{w}^{*}(v)\big)$ and $\rho_{\diff}^{*}(u, v)=\Cov\big(Z_{\diff}^{*}(u), Z_{\diff}^{*}(v)\big)$. We present explicit formulas of $\rho_{w}^{*}(u, v)$ and $\rho_{\diff}^{*}(u, v)$ in Theorem \ref{thm:cov} with the proof in Supplement S2.

\begin{theorem}
The exact expressions for $\rho_{\diff}^{*}(u, v)$ and $\rho_{w}^{*}(u, v)$ are
$$
\begin{aligned}
\rho_{w}^{*}(u, v) &=\frac{(u \wedge v)(1-(u \vee v))}{(u \vee v)(1-(u \wedge v))} \,,\\
\rho_{\diff}^{*}(u, v) &=\frac{(u \wedge v)(1-(u \vee v))}{\sqrt{(u \wedge v)(1-(u \wedge v))(u \vee v)(1-(u \vee v))}} \,,
\end{aligned}
$$
where $u \wedge v=\min (u, v)$ and $u \vee v=\max (u, v)$.
\label{thm:cov}
\end{theorem}
Theorems \ref{thm:basis} and \ref{thm:cov} together show that when $\R$ satisfies Conditions (1)-(6), the limiting distributions of \eqref{eq:imit1} and \eqref{eq:imit2}, and statistics based on \eqref{eq:imit1} and \eqref{eq:imit2}, are independent from $\R$, thus asymptotically distribution-free.

\subsection{Tail probabilities}

\allowdisplaybreaks
Based on Theorems \ref{thm:basis} and \ref{thm:cov}, we are able to approximate the tail probabilities  using Woodroofe’s method \cite{woodroofe1976frequentist,woodroofe1978large} and Siegmund’s method \cite{siegmund1988approximate,siegmund1992tail}. Specifically, following the routine of \cite{chu2019asymptotic}, and the proof of Proposition 3.4 in \cite{chen2015graph}, we can approximate the tail probabilities by 
\begin{align}
    & \Pp\big(  \max _{n_{0} \leq t \leq n_{1}} M_R(t) >b\big) \label{eq:MS}  \\
    & \approx 1-  \Pp\big(\max _{n_{0} \leq t \leq n_{1}} Z_{w}(t)<b\big) \Pp\big(\max _{n_{0} \leq t \leq n_{1}}|Z_{\diff}(t)|<b\big), \nonumber  \\
    &  \Pp\big(  \max_{1 \leq t_{1}<t_{2}\leq n \atop l_{0} \leq t_{2}-t_{1} \leq l_{1}} M_R(t_1,t_2)   >b \big)    \label{eq:MI}   \\
    & \approx  1 -   \Pp\big(\max _{l_{0} \leq t_{2}-t_{1} \leq l_{1}} Z_{w}(t_{1}, t_{2})<b\big) \nonumber \\
    & \hspace{10mm} \times\Pp\big(\max _{l_{0} \leq t_{2}-t_{1} \leq l_{1}}|Z_{\diff}(t_1, t_{2})|<b\big), \nonumber
\end{align}    
where 
\begin{align}
& \Pp\big(\max _{n_{0} \leq t \leq n_{1}} Z_{w}(t)>b\big) \label{eq:Zw Zdiff} \\
&\approx b \phi(b) \int_{\frac{n_{0}}{n}}^{\frac{n_{1}}{n}} h_{w}(n, x) \nu \big(b \sqrt{2 h_{w}(n, x) / n}\big) {\rm d} x,  \nonumber  \\
&\Pp\big(\max _{l_{0} \leq t_{2}-t_{1} \leq l_{1}} Z_{w}(t_{1}, t_{2})>b\big) \\
& \approx b^{3} \phi(b) \int_{\frac{l_{0}}{n}}^{\frac{l_{1}}{n}} \Big(h_{w}(n, x) \nu\big(b \sqrt{2 h_{w}(n, x) / n}\big)\Big)^{2}(1-x) {\rm d} x,  \nonumber  \\
&\Pp\big(\max _{n_{0} \leq t \leq n_{1}} Z_{\diff}(t)>b\big)  \\ 
& \approx b \phi(b) \int_{\frac{n_0}{n}}^{\frac{n_{1}}{n}} h_{\diff}(n, x) \nu \big(b \sqrt{2 h_{\diff}(n, x) / n}\big) {\rm d} x,   \nonumber \\
&\Pp\big(\max _{l_{0} \leq t_{2}-t_{1} \leq l_{1}} Z_{\diff}(t_{1}, t_{2})>b\big) \label{eq:Zw Zdiff2} \\
& \approx b^{3} \phi(b) \int_{\frac{l_{0}}{n}}^{\frac{l_{1}}{n}} \Big(h_{\diff}(n, x) \nu \big(b \sqrt{2 h_{\diff}(n, x) / n}\big)\Big)^{2}(1-x) {\rm d} x. \nonumber 
\end{align}
with  
$$
\begin{aligned}
 h_{w}(n, x)& =\frac{(n-1)(2 n x^{2} - 2 n x+1)}{2 x(1-x)(nx-1)(nx-n+1)},\\
 h_{\diff}(n, x) & = \frac{1}{2 x(1-x)}.    
\end{aligned}
$$
Here $v(x)$ is approximated as 
$v(x) \approx \frac{(2 / x)(\Phi(x / 2)-0.5)}{ (x / 2) \Phi(x / 2)+\phi(x / 2)}$ \cite{siegmund2007statistics}, where $\Phi(\cdot)$ and $\phi(\cdot)$ are the standard normal cumulative density function and standard normal density function, respectively. 

\subsection{Skewness correction}
\begin{figure*}[t]
  \centering
      \begin{tabular}{cc}
        \includegraphics[width=0.4\textwidth]{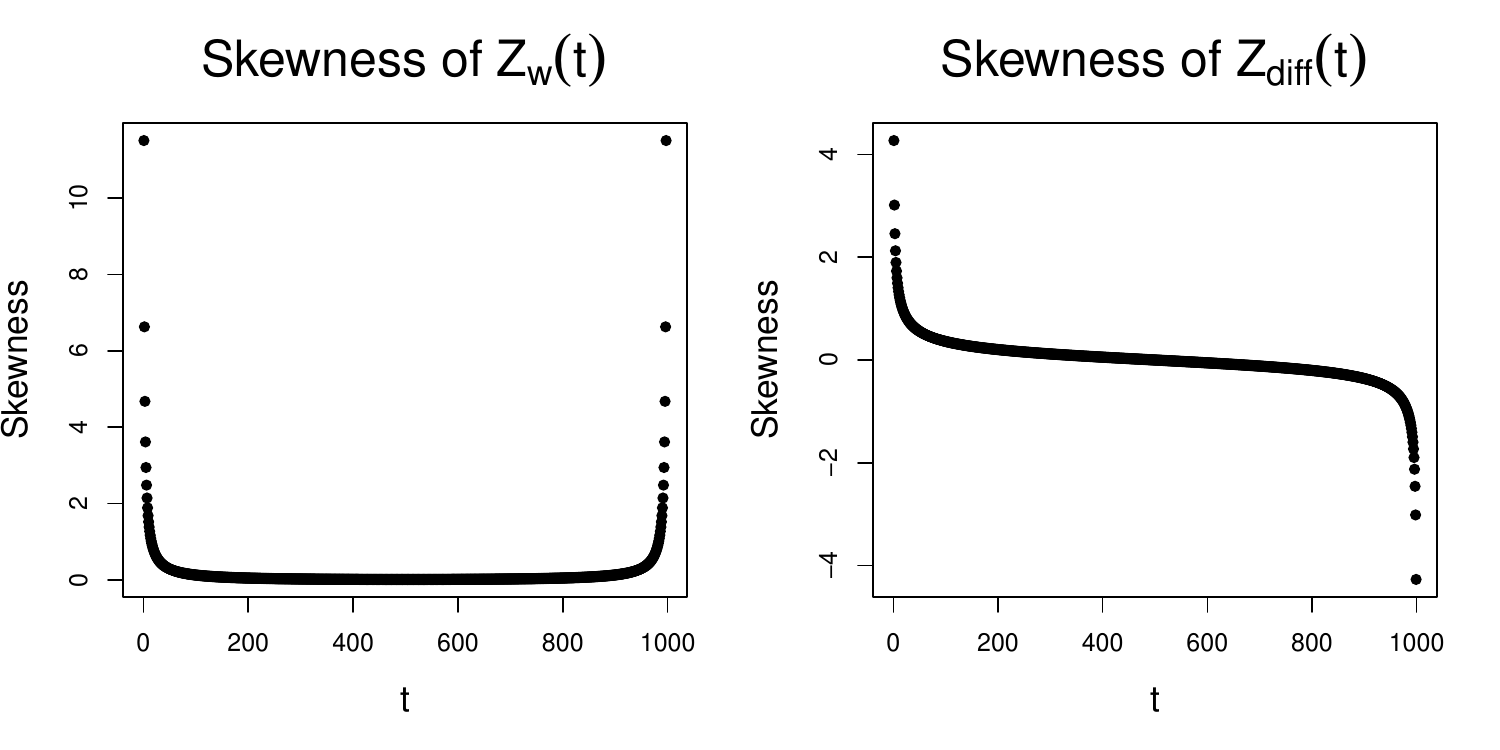}   &  \includegraphics[width=0.4\textwidth]{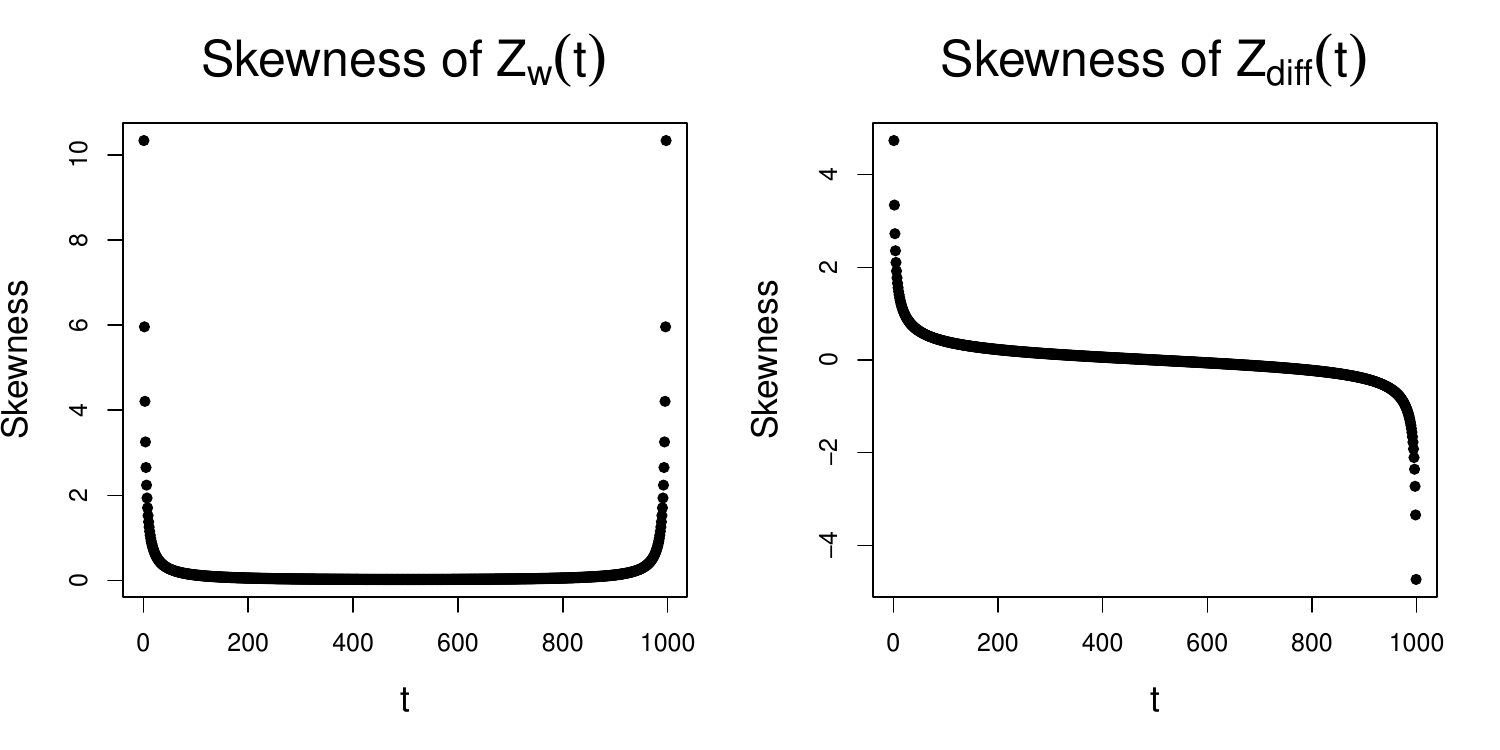} \\
    (a) $y_i \stackrel{i.i.d.}{\sim}  N_{100}(\bzero_{100},\I_{100})$ & (b) $y_i \stackrel{i.i.d.}{\sim}  Mt_{5}(\bzero_{100},\I_{100})$
      \end{tabular}
      \caption{Plots of skewness $\gamma_j(t) = \Ep\big( Z_j^3(t) \big), j = w, \diff$ against $t$ with the graph-induced rank in $10$-NNG constructed on Euclidean distance on a sequence of $1000$ points.}
\label{fig:skewness}
\end{figure*}

As observed by \cite{chen2015graph, chu2019asymptotic}, the analytical approximations of \eqref{eq:MS} and \eqref{eq:MI} can be improved by skewness correction when $n_0$ and $n-n_1$ decrease. It can be seen clearly in Figure~\ref{fig:skewness} that $Z_w(t)$  and $Z_\diff(t)$  are more skewed toward the two ends.
To be specific, instead of using \eqref{eq:Zw Zdiff}-\eqref{eq:Zw Zdiff2} to approximate \eqref{eq:MS} and \eqref{eq:MI}, we use 
\begin{align}
&\Pp\big(\max _{n_{0} \leq t \leq n_{1}} Z_{w}(t)>b\big) \label{eq:Zcorr} \\ 
&\approx b \phi(b)  \int_{\frac{n_{0}}{n}}^{\frac{n_{1}}{n}} K_{w}(n x) h_{w}(n, x) \nu \big(b \sqrt{2 h_{w}(n, x) / n}\big) {\rm d} x,  \nonumber\\
&\Pp\big(\max _{l_{0} \leq t_{2}-t_{1} \leq l_{1}} Z_{w}(t_{1}, t_{2})>b\big)   \\ 
& \approx b^{3} \phi(b)  \int_{\frac{l_{0}}{n}}^{\frac{l_{1}}{n}}  \Big(h_{w}(n, x) \nu\big(b \sqrt{2 h_{w}(n, x) / n}\big)\Big)^{2} \nonumber \\
& \hspace{25mm} \times K_{w}(n x)(1-x) {\rm d} x, \nonumber \\
&\Pp\big(\max _{n_{0} \leq t \leq n_{1}} Z_{\diff}(t)>b\big) \label{eq:Zcorr2}  \\
&  \approx b \phi(b) \int_{\frac{n}{n}}^{\frac{n_{1}}{n}} K_{\diff}(n x) h_{\diff}(n, x) \nu \big(b \sqrt{2 h_{\diff}(n, x) / n}\big) {\rm d} x,  \nonumber \\
&\Pp\big(\max _{l_{0} \leq t_{2}-t_{1} \leq l_{1}} Z_{\diff}(t_{1}, t_{2})>b\big)   \\
& \approx b^{3} \phi(b) \int_{\frac{l_{0}}{n}}^{\frac{l_{1}}{n}}  \Big(h_{\diff}(n, x) \nu \big(b \sqrt{2 h_{\diff}(n, x) / n}\big)\Big)^{2} \nonumber \\
& \hspace{25mm} \times K_{\diff}(n x)(1-x) {\rm d} x,  \nonumber
\end{align}
where 
$K_{j}(t)=\frac{\exp \Big(\frac{1}{2}\big(b-\hat{\theta}_{b, j}(t)\big)^{2}+\frac{1}{6} \gamma_{j}(t) \hat{\theta}_{b, j}(t)^{3}\Big)}{\sqrt{1+\gamma_{j}(t) \hat{\theta}_{b, j}(t)}}$ for $j = w,\diff$, with $\hat{\theta}_{b, j}(t) = \frac{-1+\sqrt{1+2 \gamma_{j}(t) b}}{\gamma_{j}(t)}$ and  $\gamma_{j}(t)=\Ep\big(Z_{j}^{3}(t)\big)$. The only unknown quantities in the above expressions are $\gamma_{w}(t)$ and $\gamma_{\diff}(t)$, whose exact analytic expressions are quite long and are thus provided in  Supplement S3. 

\subsection{Assessment of finite sample approximations}
\label{sec:assess}
\begin{figure}[t] 
  \centering
 \includegraphics[width=0.49\textwidth]{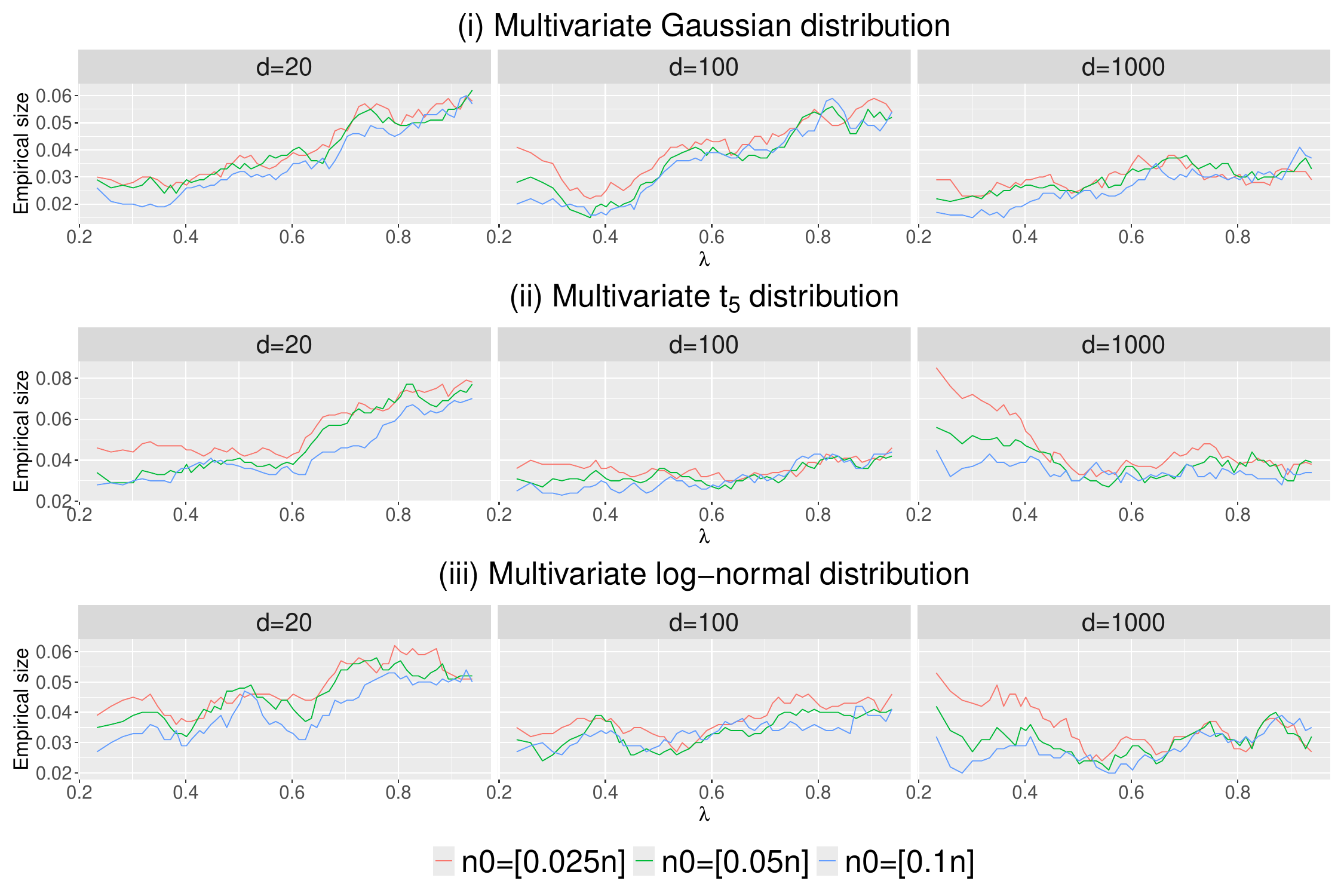}
 \caption{Empirical size of $\Mg$-NN after skewness correction at $0.05$ nominal level with $n = 1000$ under settings (i), (ii) and (iii). The $k$-NNG for $k = [n^{\lambda}]$ is considered.
 }
\label{fig:size} 
\end{figure}

Here we assess the performance of the asymptotic approximations with finite samples. For a constant $a$, we define the first-order auto-regressive correlation matrix as $\bSigma(a) =  (a^{|i-j|})_{i,j=1}^d \in \RR^{d \times d}$. We consider three distributions for three different dimensions $d = 20, 100$ and $1000$ with $n=1000$:
\begin{enumerate}
    \item[(i)] Multivariate Gaussian:  $y_i \sim N_d\big(\bzero_d,\bSigma(0.6)\big)$;
    \item[(ii)] Multivariate $t_5$:  $y_i \sim Mt_5\big(\bzero_d,\bSigma(0.5)\big)$;
    \item[(iii)] Multivariate log-normal:  $y_i \sim \exp\big( N_d\big(\bzero_d,\bSigma(0.4)\big) \big)$.
\end{enumerate}
We report the empirical sizes estimated by $1,000$ Monte Carlo simulations. 
Here, we focus on the graph-induced rank in $k$-NNG. We denote the 
scan statistic $M_R(t)$ on the graph-induced rank in $k$-NNG by $\Mg$-NN. We set $n_1 = n - n_0$ and consider $n_0= [0.025n], [0.05n], [0.1n]$. The nominal level is set to be $0.05$. We see that the empirical sizes are well controlled across all settings even for $n_0$ as small as $[0.025n]$ (Table \ref{tabp2}). We further present the empirical sizes of the $\Mg$-NN in Figure~\ref{fig:size}, considering $k = [n^{\lambda}]$ with $\lambda$ ranging from $0.2$ to  $0.95$ in $k$-NNG. This visualization indicates that the empirical sizes are close to the nominal levels across a broad spectrum of $\lambda$ values.

\begin{table*}%[bh]%[htb!]
    \centering \caption{Empirical size of $\Mg$-NN after skewness correction at $0.05$ nominal level with $n = 1000$ under settings (i), (ii) and (iii). The $k$-NNG for various $k$'s is considered. Here $k_1 = [n^{0.5}]$, $k_2 = [n^{0.65}]$ and $k_3 = [n^{0.8}]$.}
    \label{tabp2} 
 \begin{tabular}{|c|l|ccc|ccc|ccc|}
\hline
\multirow{2}{*}{Setting}& &  \multicolumn{3}{c|}{$n_0 = [0.1n]$} &  \multicolumn{3}{c|}{$n_0 = [0.05n]$} & \multicolumn{3}{c|}{$n_0 = [0.025n]$} \\
& \diagbox{$k$}{$d$} & $20$ & $100$ & $1000$ & $20$ & $100$ & $1000$ & $20$ & $100$ & $1000$\\
\hline \hline 
\multirow{5}{*}{(i)} & $5$ & 0.04 & 0.02 & 0.02 & 0.04 & 0.03 & 0.02 & 0.05 & 0.04 & 0.04\\
& $10$ & 0.03 & 0.02 & 0.03 & 0.04 & 0.03 & 0.03 & 0.05 & 0.03 & 0.04\\
& $k_1$ & 0.03 & 0.03 & 0.03 & 0.03 & 0.03 & 0.03 & 0.04 & 0.03 & 0.03\\
& $k_2$ & 0.03 & 0.03 & 0.03 & 0.04 & 0.03 & 0.03 & 0.04 & 0.03 & 0.03\\
& $k_3$ & 0.04 & 0.03 & 0.04 & 0.05 & 0.04 & 0.04 & 0.06 & 0.04 & 0.03\\ 
\hline
\hline
\multirow{5}{*}{(ii)} &$5$ & 0.03 & 0.03 & 0.04 & 0.02 & 0.03 & 0.06 & 0.04 & 0.04 & 0.08\\
& $10$ & 0.03 & 0.03 & 0.04 & 0.03 & 0.03 & 0.04 & 0.04 & 0.04 & 0.06\\
& $k_1$ & 0.04 & 0.03 & 0.03 & 0.04 & 0.03 & 0.03 & 0.04 & 0.04 & 0.03\\
& $k_2$ & 0.04 & 0.04 & 0.03 & 0.05 & 0.03 & 0.03 & 0.05 & 0.04 & 0.04\\
& $k_3$ & 0.06 & 0.05 & 0.03 & 0.06 & 0.05 & 0.03 & 0.07 & 0.05 & 0.03\\
\hline
\hline
\multirow{5}{*}{(iii)} & $5$ & 0.03 & 0.03 & 0.03 & 0.03 & 0.04 & 0.05 & 0.05 & 0.05 & 0.06\\
& $10$ & 0.03 & 0.04 & 0.03 & 0.03 & 0.04 & 0.04 & 0.05 & 0.04 & 0.05\\
& $k_1$ & 0.04 & 0.03 & 0.02 & 0.05 & 0.03 & 0.02 & 0.05 & 0.03 & 0.03\\
& $k_2$ & 0.04 & 0.03 & 0.02 & 0.05 & 0.03 & 0.02 & 0.06 & 0.04 & 0.03\\
& $k_3$ & 0.05 & 0.04 & 0.03 & 0.06 & 0.03 & 0.03 & 0.07 & 0.04 & 0.03\\
\hline
\end{tabular}
\end{table*}

\subsection{Consistency}

\label{sec:consistency}

We here examine the consistency of $M_R$ when $k$-NNG or $k$-MST is used to get the graph-induced rank. At first, we define the limits 
$$
\begin{aligned}
& M(\delta_1, \delta_2 ) =  \lim_{n \rightarrow \infty } \frac{M_R\big(  [\delta_1 n], [\delta_2 n] \big)}{ \sqrt{ n}} \text{ for } 0 \leq \delta_1 < \delta_2 < 1 
\end{aligned}
$$ 
and $M(\delta) = M(0, \delta)$ for $0 < \delta < 1$. 
\begin{theorem}
Consider two continuous multivariate distributions $F_0$ and $F_1$ which differ on a set of positive Lebesgue measures and the graph-induced rank is used with $k$-MST or $k$-NNG based on the Euclidean distance, where $k = O(1)$. 
\begin{itemize}
    \item For the change-point alternative $H_1$: let $\omega= \lim_{n \rightarrow \infty } \tau/n \in (0,1)$,  and $\hat \omega = \hat \tau/n$. Assume that
\begin{equation}
\sup_{\delta \in (0,1) } \Big|\frac{M_R([\delta n])}{\sqrt{ n}} - M(\delta) \Big| \stackrel{P}{\rightarrow} 0  
\label{asump:limiting1}
\end{equation}    
as  $n \rightarrow \infty$, where $\stackrel{P}{\rightarrow}$ denotes the convergence in probability.  Then the scan statistic of $M_R(t)$ is consistent in that it will reject $H_0$ against $H_1$ with probability going to one for any significance level $0 < \alpha < 1$ and 
$$\Pb\big( |\hat \omega - \omega| > \epsilon \big) \rightarrow 0  \text{ as } n \rightarrow \infty \text{ for any } \epsilon > 0 \,.$$    
    \item For the changed interval alternative $H_2$: let  $\omega_i = \lim_{n \rightarrow \infty } \tau_i/n \in (0,1)$, and $\hat \omega_{i} = \hat \tau_{i}/n$ for $i=1,2$. Assume that  $ \omega_2 - \omega_1 > 0$ and  
    \begin{equation}
      \sup_{0 < \delta_1 < \delta_2 <1 } \Big|\frac{M_R([\delta_1 n], [\delta_2 n])}{\sqrt{ n}} - M(\delta_1,\delta_2) \Big| \stackrel{P}{\rightarrow} 0 
   \label{asump:limiting2}
    \end{equation}
     as  $n \rightarrow \infty$, then the scan statistic of  $M_R(t_1,t_2)$ is consistent in that it will reject $H_0$ against $H_2$ with probability going to one for any significance level $0 < \alpha < 1$ and 
$$\Pb\big( \cup_{i=1}^2 \{ |\hat \omega_i - \omega_i| > \epsilon \} \big) \rightarrow 0 \text{ as } n \rightarrow \infty \text{ for any } \epsilon > 0 \,.$$    
\end{itemize}
\label{thm:con 1}
\end{theorem}

\begin{figure*}[!b] 
  \centering
  \includegraphics[width=0.75\textwidth]{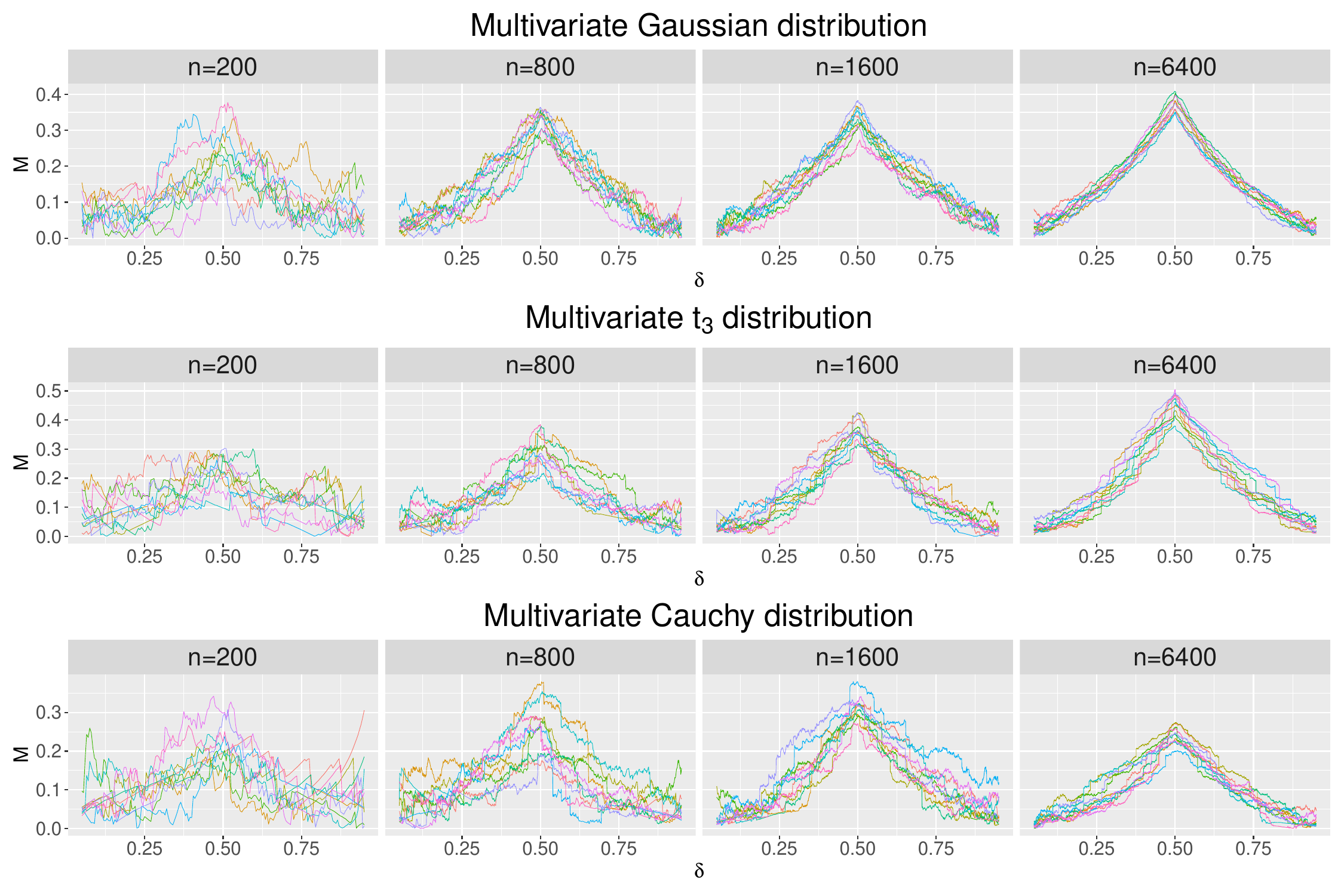}
 \caption{ \label{fig:limit2}  Ten independent sequences (depicted using different colors) of $M_R( [\delta n])/\sqrt{n}$ against $\delta$ for $n = 200, 800, 1600$ and $6400$ for the three settings.} 
\end{figure*}
 
The proof of this theorem is provided in Supplement S4. Although Assumptions \eqref{asump:limiting1} and \eqref{asump:limiting2} are reasonable, their verification is difficult and is left for future work. Here we check them numerically through Monte Carlo simulations. Specifically, we consider the following combinations of $(F_0,F_1)$ with $\omega = 0.5$ and $d=500$: 
\begin{enumerate}
    \item Multivariate Gaussian: $\big( N_{d}( \bzero_d, \I_d), N_{d}( 0.1 \bone_d, \I_d) \big)$;
    \item Multivariate $t_3$: $\big( Mt_{3}( \bzero_d, \I_d), Mt_{3}( 0.1 \bone_d, 1.02^2 \I_d \big)$;
    \item Multivariate Cauchy: $\big( MC_d( \bzero_d, \I_d), MC_d(2 \bone_d, \I_d) \big)$. 
\end{enumerate}
We generate $10$ independent sequences for each setting and the plots of $M_R( [\delta n])/\sqrt{n}$ against $\delta$ for various values of $n$ are presented Figure \ref{fig:limit2}. These plots verify the assumption that  $M_R( [\delta n])/\sqrt{n}$ converges when $n \rightarrow \infty$.

\begin{remark}
The theoretical results in Theorem \ref{thm:con 1} are derived under the assumption of $k = O(1)$. However, in practice, we have observed that using larger values of $k$, specifically $k = O(n^\lambda)$ for $\lambda > 0$, leads to increased power. This improvement can be attributed to the fact that larger $k$ values incorporate more similarity information into the graph. Although increasing $k$ may introduce additional noise, when $\lambda$ is not too large, the benefits of additional similarity information outweigh the noise, compared to the case where $k=O(1)$. 
The proof of Theorem \ref{thm:con 1} builds on the foundational work by \cite{henze1999multivariate}, which considered the minimum spanning tree, i.e., $k=1$. If a similar result to that in \cite{henze1999multivariate} can be established for $k = O(n^\lambda)$ with $\lambda > 0$, we believe that the assumption in Theorem \ref{thm:con 1} could be relaxed to $k = O(n^\lambda)$ as well. We leave this extension for future research.
\end{remark}

\section{Simulation studies}
\label{sec: simulation}

\subsection{The choice of k}
\label{sec:choice}

The choice of graphs remains an open question for CPD based on similarity graphs \cite{chen2015graph, chu2019asymptotic, zhang2021graph}. We adapt the method in \cite{zhang2021graph} and \cite{rise}. Specifically, they compare the empirical power of the method for different choices of $k = [n^{\lambda}]$ by varying $\lambda$ from $0$ to $1$. \cite{zhang2021graph} suggested using $k = [n^{0.5}]$ for the generalized edge-count test (GET) \cite{chu2019asymptotic} when the $k$-MST is used, while \cite{rise} recommended $k = [n^{0.65}]$ for their two-sample test statistic with graph-induced rank on the $k$-NNG. We follow the same way in choosing $k$ in $\Mg$-NN.  We generate independent sequences from three different distribution pairs of $(F_0,F_1)$: 
\begin{enumerate}
    \item Multivariate Gaussian: $\big( N_{d}( \bzero_d, \I_d), N_{d}( \frac{30}{\sqrt{n d}} \bone_d, \I_d) \big)$;
    \item Multivariate $t_3$: $\big(Mt_{3}( \bzero_d, \I_d), Mt_{3}( \frac{30}{\sqrt{ n d}} \bone_d, (1 + \frac{30}{\sqrt{n d}})^2 \I_d) \big)$;
    \item Multivariate Cauchy: $\big( MC_d( \bzero_d, \I_d), MC_d( \frac{30}{\sqrt{n}} \bone_d, \I_d) \big)$. 
\end{enumerate}
The parameters are chosen so that the tests have moderate power. The change-point is set to be $\tau = n/2$, the dimension $d = 500$ and $n = 50, 100, 200$. We set $n_0 = \lceil 0.05 n \rceil$ and $n_1 = n - n_0$, which will also be our choice by default in latter experiments, where $\lceil x \rceil$ denotes the smallest integer larger than or equal to $x$. For comparison, we also include two graph-based methods, GET and the max-type edge-count test (MET) proposed in \cite{chu2019asymptotic} using the $k$-MST. The empirical power is defined as the ratio of successful detection where the $p$-value is smaller than $0.05$. For fairness, the $p$-values are approximated by $1,000$ random permutations for all methods. 

\begin{figure}[!t] 
  \centering
 \includegraphics[width=250pt]{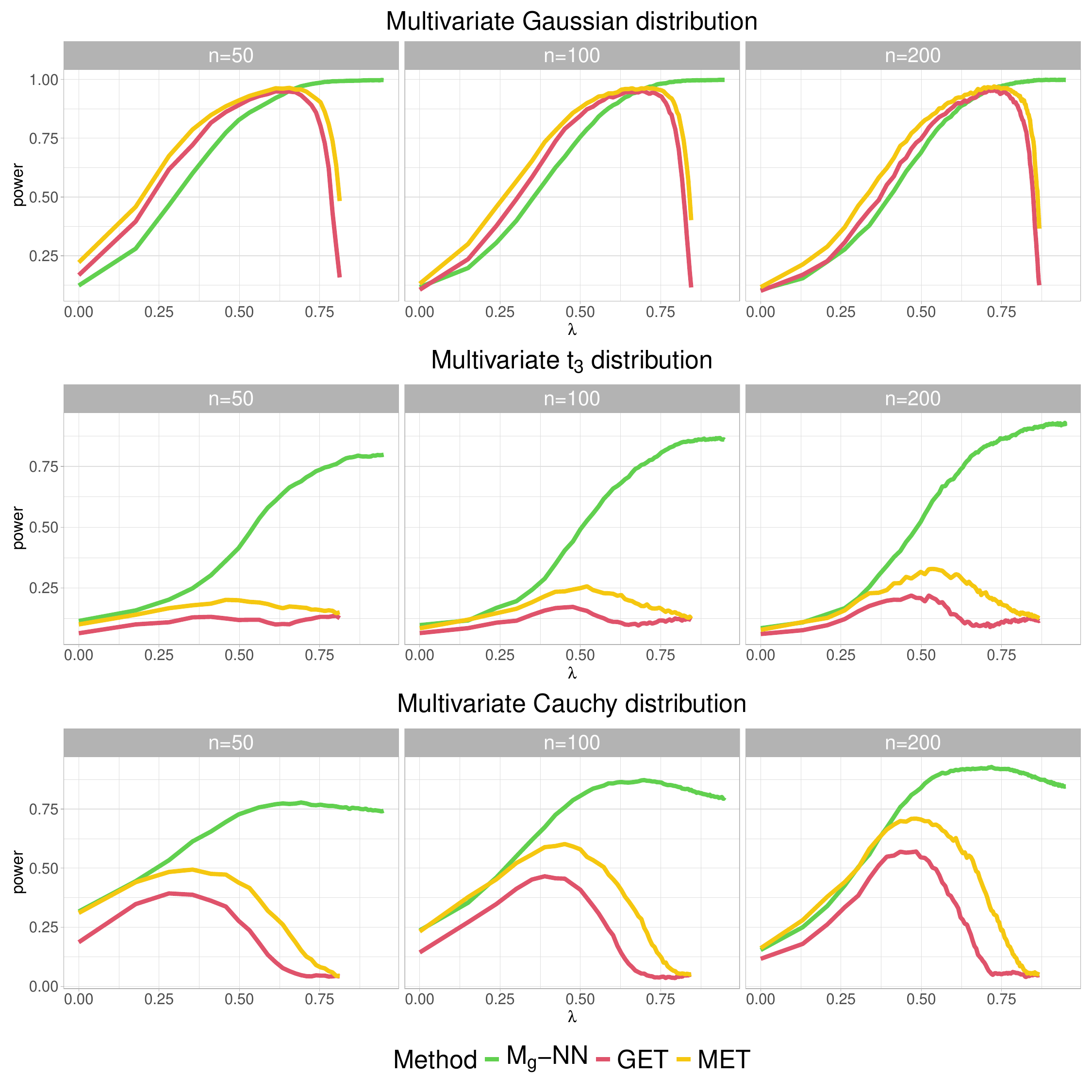}
 \caption{Empirical power of $\Mg$-NN, GET,  and MET over $1000$ times of repetitions under each setting.
 }
\label{fig:power1} 
\end{figure}

Figure~\ref{fig:power1} illustrates the power of the tests for different values of $k = \lfloor n^{\lambda} \rfloor$.  For $\Mg$-NN, $\lambda \in (0, 0.95]$, while for GET and MET, $k$ is chosen from $1$ to $\min \{\lfloor n^{0.95} \rfloor, \lfloor \frac{n-1}{2} \rfloor \}$.
This range is chosen because the $l$th-MST contains $n-1$ edges, whereas a complete graph has $n(n-1)/2$ edges. We observe that the power of these tests initially increases rapidly as $k$ or $\lambda$ increases. However, as $k$ continues to increase, the power of GET and MET decreases sharply.  In contrast, $\Mg$-NN demonstrates greater robustness, maintaining consistent performance even for larger values of $k$ (e.g., $k > n^{0.65}$). In the case of the multivariate Cauchy distribution, the power of $\Mg$-NN exhibits a slight decrease when $k > n^{0.65}$, but overall remains stable. This robustness of $\Mg$-NN can be attributed to its use of ranks on the edges, which effectively reduces the impact of additional noise from denser graphs by assigning smaller ranks to edges introduced later. As a result, $\Mg$-NN consistently achieves the best performance, showing a significant improvement in power for heavy-tailed distributions (such as the multivariate $t_3$ and Cauchy distributions) and demonstrating resilience across a wide range of $k$ values. Considering the trade-offs between computational cost, control of type I error, and power, we select $\lambda = 0.65$ for $\Mg$-NN, 
and $\lambda = 0.5$ for GET and MET in subsequent analyses. These choices strike a reasonable balance and are consistent with previous recommendations \cite{zhang2021graph, rise}. 

\begin{remark}
\label{adhoc}
Selecting an appropriate value of $k$ remains an open question for graph-based methods in real-world applications. Beyond using a fixed value, such as $k=[n^{0.65}]$, we propose a data-driven approach to empirically determine $k$. Specifically, we first randomly permute the observations, then apply a mean shift to the last  $\lceil n/2 \rceil$ observations by adding a value $\mu$ to each dimension. We then select the $k$ value that produces the smallest $p$-value. This approach incorporates the unknown underlying distribution for  choosing an optimal value of $k$.
\end{remark}

\subsection{Performance comparison}

We compare RING-CPD to GET and MET on $k$-MST 
with 
$k = [n^{0.5}]$, the graph-based method on the shortest Hamiltonian path \cite{shi2017consistent} (SWR), the method based on Fr\'echet means and variances \cite{dubey2020frechet} (DM). We also compare three interpoint distance-based methods, the widely used distance-based method E-Divisive (ED) \cite{matteson2014nonparametric} implemented in the R package \textit{ecp}, and the other two methods proposed recently by \cite{li2020asymptotic} and \cite{nie2021weighted}. \cite{li2020asymptotic} proposed four statistics and we compare the statistic $C_{2N}$ that had a satisfactory performance in most of their simulation settings. \cite{nie2021weighted} proposed three statistics, which perform well for location change, scale change, and general change, respectively. Here we compare with their statistic $S_3$, which they concluded to have relatively robust performance across various alternatives.  For fairness, the $p$-values of all these methods are approximated by $1,000$ random permutations. 

\begin{table*}[!b]
\small
    \centering
      \caption{The specific changes for different settings and alternatives.  }
    \begin{tabular}{|c|c| c|c| c| c c| cc|}
    \hline %Setting/Alternative
    & & \multicolumn{7}{c|}{ Alternative}\\  \cline{3-9}
\multirow{2}{*}{ Setting }    & $H_0$ & (a) & (b) & (c) & \multicolumn{2}{c|}{(d)} & \multicolumn{2}{c|}{ (e) } \\ 
      &   $\bSigma_0$ & $\delta$ & $\sigma$ & $\bSigma_1$ & $\delta$ & $\sigma$ & $\delta$ &  $\bSigma_1$ \\ \hline 
        
 (I) & $\bSigma(0.6)$ &  $ \frac{2 \log d}{5 \sqrt{d}}$ & $\sqrt{ \frac{\log d}{16 d}}$ & $\bSigma(0.16)$ & $ \frac{ \log d}{10 \sqrt{d}}$ & $ \sqrt{ \frac{\log d}{16 d}}$ & $ \sqrt{ \frac{\log d}{4 d}}$ & $\bSigma(0.3)$\\
 
  (II) & $\bSigma(0.6)$ &  $ \frac{5 \log d}{4 \sqrt{d}}$ & $ \frac{3 \log d}{10 \sqrt{ d}}$ & $0.6 \bSigma(0.1)$ & $ \frac{ \log d}{3 \sqrt{d}}$ & $ \frac{3 \log d}{10  \sqrt{ d}}$ & $ \frac{\log d}{2 \sqrt{ d}}$ & $\bSigma(0.8)$\\
  
   (III) & $\bSigma(0.4)$ &  $ \frac{11 \log d}{20 \sqrt{d}}$ & $ \frac{6 \log d}{5 d^{2/5}}$ & $ \bSigma(0.85)$ & $ \frac{6 \log d}{25 d^{2/5}}$ & $ \sqrt{ \frac{\log d}{25  d}}$ & $ \frac{6 \log d}{25 d^{2/5}}$ & $\bSigma(0.6)$\\
   
    (IV) & $\A_0$ &  $ \frac{5 \log d}{2 \sqrt{d}}$ & $ \frac{ 9}{ 10 \sqrt{d}}$ &  $\A_1$ & $ \sqrt{\frac{49 \log d}{16 d}}$ & $ \frac{3}{4  \sqrt{ d}}$ &$ \sqrt{\frac{49 \log d}{16 d}}$& $\A_2$\\
    
     (V) & $\I_d$ &   $ \frac{3}{5 \log d}$ & $\sqrt{ \frac{ \log d}{ 25 d}}$  & $\bSigma(0.55)$ & $\frac{3}{10  \log d}$ & $  \sqrt{ \frac{\log d}{25 d}}$ & $\frac{3}{10  \log d}$  & $\bSigma(0.48)$\\
     
      (VI) & $\bSigma(0.5)$ &  $ \frac{7 \log d}{20 \sqrt{d}}$ & $ \frac{\log d}{5 \sqrt{ d}}$ & $ \bSigma(0.1)$ & $ \frac{ \log d}{5 \sqrt{d}}$ & $ \frac{\log d}{5  \sqrt{ d}}$ & $ \frac{\log d}{5 \sqrt{ d}}$ & $\bSigma(0.15)$\\
\hline
    \end{tabular}
    \label{tab:setting}
\end{table*}

 We set $n = 200$ and the change-point $\tau = [n/3]$ and consider the dimension of the distributions $d = 200, 500, 1000$. Before the change-point $y_i \sim F_0$ and after the change-point $y_i \sim F_1$. We consider both the empirical power and the detection accuracy estimated from $1000$ trials for each scenario. The empirical power is the ratio of the successful detection defined as $p$-value smaller than the nominal level $0.05$. The detection accuracy is provided in parentheses, which is the ratio of trials that the detected change-point is located in $[\tau - 0.05n, \tau + 0.05n]$ and the $p$-value smaller than $0.05$. We consider various settings that cover light-tailed, heavy-tailed, skewed, and mixture distributions for location, scale, and mixed alternatives. Specifically, we consider six settings for $F_i$, $i=0,1$: 
 \begin{enumerate}
     \item[(I)] Multivariate Gaussian: $N_{d}(\bmu_i, \bSigma_i )$;
     \item[(II)] Multivariate  $t_5$: $Mt_{5}(\bmu_i, \bSigma_i)$;
     \item[(III)] Multivariate Cauchy: $MC_d(\bmu_i, \bSigma_i)$;
     \item[(IV)] Multivariate $\chi_5^2$: $\chi_5^2(\bmu_i, \bSigma_i)$ (generated as $\bSigma_i^{\frac{1}{2}} (X - 5\bone_d + \bmu_i)$ where the $d$ components of $X$ are i.i.d. $\chi_5^2$);
     \item[(V)] Multivariate Gaussian mixture:  $W N_d(\bmu_i,\bSigma_i ) + (1-W) N_d(-\bmu_i,\bSigma_i)$ with $W \sim {\rm Bernoulli}(0.5)$; 
     \item[(VI)] Multivariate Gaussian with Multivariate $t_7$ outliers $W N_d(\bmu_i,\bSigma_i) + (1-W) Mt_7\big(\bmu_i,\bSigma_i)$ with $W \sim {\rm Bernoulli}(0.9)$. 
 \end{enumerate}
 %(I) the multivariate Gaussian distribution $N_{d}(\bmu_i, \bSigma_i )$; (II)  (III)  (IV)  (V)  and (VI) 
 We set $\bmu_0 = \bzero_d$ for $F_0$ and $\bmu_1 = \delta \bone_d$ for $F_1$, where $\delta$ is different for different settings. For each setting, we consider five different changes: \begin{enumerate}
     \item[(a)] location ($\delta \neq 0$ and $\bSigma_1 = \bSigma_0$);
     \item[(b)] simple scale ($\delta = 0$ and $\bSigma_1 = (1+\sigma)^2 \bSigma_0$); 
     \item[(c)] complex scale ($\delta = 0$ and $\bSigma_1 \neq \bSigma_0$);
     \item[(d)] location and simple scale mixed ($\delta \neq 0$ and $\bSigma_1 = (1+\sigma)^2 \bSigma_0$); 
     \item[(e)] location and complex scale mixed ($\delta \neq 0$ and $\bSigma_1 \neq \bSigma_0$). 
 \end{enumerate}
  The choice of $\delta$, $\sigma$, and $\bSigma_i$, $i = 1, 2$ are specified differently for the settings and alternatives, summarized in  Table~\ref{tab:setting}, where the changes in signal are set so that the best test has moderate power to be comparable. Here for Setting IV, the covariance matrices $\A_i = \V \B_i \V$, for $i = 0,1,2$, where $\V$ is a diagonal matrix with the diagonal elements sampled independently from ${\rm U}(1,3)$, $\B_i = {\rm diag}(\B_{i 1}, \ldots, \B_{i \frac{d}{10}})$ is a block-diagonal correlation matrix. Each diagonal block $\B_{i j}$ is a $10 \times 10$ matrix with diagonal entries being $1$ and off-diagonal entries equal to $\rho_{i j} \sim {\rm U}(a_j,b_j)$ independently.  We set $a_0 = 0, b_0 = 0.5$,  $a_1 = 0.3, b_1 = 0.8$ and $a_3 = 0.2, b_3 = 0.7$. Each configuration is repeated $1,000$ times. We present the results of Settings I-III in Table \ref{tab p1},  and the results of Settings IV-VI in Table \ref{tab p2}. Under each setting, the largest value and those larger than $95\%$ of the largest value are highlighted in bold.

 \begin{table*}[!htp]
 \small
    \centering
 \caption{The empirical power (detection accuracy) in percent
 under Settings I-III. { Here $\Mg$-NN is the proposed method.}
 }
    \label{tab p1} 
       \begin{tabular}{|l|ccc|ccc|ccc|}
\hline
 & \multicolumn{3}{c|}{Setting I (Gaussian)} & \multicolumn{3}{c|}{Setting II ($t_5$)}  & \multicolumn{3}{c|}{Setting III (Cauchy) }\\ 
$d$ & 200 & 500 & 1000 & 200 & 500 & 1000 & 200 & 500 & 1000\\
  \hline 
 & \multicolumn{9}{c|}{  (a) Location change }  \\
\hline 
$\Mg$-NN & 76(58) & 67(49) & 59(40) & \textbf{89(71)} & \textbf{79(59)} & \textbf{67(49)} & \textbf{99(88)} & \textbf{91(78)} & \textbf{72(55)}\\
GET & 63(46) & 52(36) & 40(25) & 68(48) & 41(26) & 20(12) & 85(72) & 54(40) & 28(17)\\
MET & 68(50) & 58(39) & 46(30) & 75(52) & 50(32) & 31(18) & 90(75) & 67(50) & 44(26)\\
SWR & 21(8) & 18(6) & 16(4) & 19(6) & 19(6) & 15(4) & 44(23) & 40(18) & 32(15)\\
DM & 7(0) & 6(0) & 7(0) & 6(0) & 5(0) & 4(0) & 5(0) & 4(0) & 5(0)\\
ED & \textbf{97(85)} & \textbf{96(83)} & \textbf{95(80)} & 73(57) & 28(19) & 12(4) & 6(1) & 5(0) & 4(1)\\
$C_{2N}$ & \textbf{95(81)} & \textbf{93(81)} & \textbf{90}(75) & 53(34) & 19(7) & 8(2) & 5(0) & 5(0) & 6(0)\\
$S_3$ & 5(1) & 5(1) & 6(0) & 6(0) & 5(0) & 4(0) & 5(0) & 4(0) &5(0)\\
\hline
& \multicolumn{9}{c|}{(b) Simple scale change }  \\ 
\hline 
$\Mg$-NN & 65(38) & 74(46) & 80(51) & \textbf{99(78)} & \textbf{94(68)}& \textbf{82}(47) & \textbf{98(70)} & \textbf{90(56)} & \textbf{81(46)}\\
GET & 61(33) & 71(40) & 74(44) & \textbf{99(75)} & {86(56)} & {69(35)} & \textbf{97(68)} & {83(46)} & {63(32)}\\
MET & 63(36) & 72(42) & 76(47) & \textbf{99(76)} & \textbf{91}(63) & {76(42)} & \textbf{98(68)} & \textbf{90(56)} & \textbf{77(43)}\\
SWR & 5(0) & 5(0) & 5(0) & 33(14) & 19(7) & 13(3) & 23(10) & 20(5) & 12(3)\\
DM & 63(36) & 50(21) & 32(4) & 72(47) & 57(34) & 43(24) & 4(0) & 4(0) & 4(0)\\
ED & 5(2) & 6(1) & 6(1) & \textbf{98(78)} & \textbf{93(69)} & \textbf{83(56)} & 30(12) & 19(8) & 18(6)\\
$C_{2N}$ & 73(41) & 84(53) & \textbf{88}(57) & 73(42) & 66(27) & 54(11) & 5(0) & 5(0) & 4(0)\\
$S_3$ & \textbf{83(54)} & \textbf{90(64)} & \textbf{92(67)} & 66(42) & 49(28) & 37(19) & 4(0) & 4(0) & 4(0)\\
\hline
 & \multicolumn{9}{c|}{(c) Complex scale  change }  \\
\hline 
$\Mg$-NN & \textbf{96}(73) & \textbf{96}(73) & \textbf{95}(72) & \textbf{97(85)} & {85(61)} & 70(39) & \textbf{95}(80) & 86(67) & 76(55)\\
GET & 84(63) & 79(56) & 79(56) & 90(76) & 35(2) & 14(1) & \textbf{96}(84) & 77(61) & 54(38)\\
MET & 78(48) & 77(44) & 76(46) & 81(60) & 37(10) & 24(1) & 94(78) & 83(64) & 70(50)\\
SWR & 80(61) & 84(64) & 82(64) & \textbf{96(84)} & \textbf{97(84)} & \textbf{96(83)} & \textbf{99(92)} & \textbf{98(88)} & \textbf{96(84)}\\
DM & 8(0) & 6(0) & 7(0) & 70(46) & 70(43) & 68(40) & 5(0) & 5(0) & 5(0)\\
ED & 10(2) & 10(2) & 8(2) & 95(72) & 97(74) & 95(75) & 5(1) & 6(0) & 4(1)\\
$C_{2N}$ & 7(1) & 7(1) & 7(2) & 74(40) & 77(27) & 75(15) & 5(0) & 4(0) & 6(0)\\
$S_3$ & 8(1) & 9(1) & 8(1) & 67(43) & 67(41) & 66(39) & 5(0) & 5(0) & 5(0)\\
\hline
& \multicolumn{9}{c|}{ (d) Location and simple scale mixed change }  \\
\hline 
$\Mg$-NN & 69(44) & 73(46) & 80(53) & \textbf{69(48)} & \textbf{54(34)} & \textbf{37(21)} & \textbf{70(52)} & \textbf{60(43)} & \textbf{47(32)}\\
GET & 64(37) & 67(39) & 76(46) & 53(34) & 28(13) & 12(4) & 37(24) & 23(14) & 16(8)\\
MET & 66(39) & 71(43) & 77(49) & 51(30) & 28(12) & 16(5) & 49(32) & 34(20) & 28(14)\\
SWR & 5(0) & 5(0) & 5(0) & 13(3) & 12(3) & 11(3) & 20(7) & 22(8) & 19(6)\\
DM & 66(40) & 47(19) & 32(5) & 14(4) & 8(2) & 7(1) & 5(0) & 4(0) & 4(0)\\
ED & 9(2) & 9(3) & 8(2) & 60(39) & 32(18) & 19(6) & 6(1) & 5(1) & 4(1)\\
$C_{2N}$ & 77(44) & 83(54) & \textbf{89}(61) & 37(16) & 16(4) & 10(2) & 6(0) & 5(0) & 5(0)\\
$S_3$ & \textbf{84(56)} & \textbf{88(63)} & \textbf{92(69)} & 11(2) & 7(1) & 6(1) & 5(0) & 4(0) & 4(0)\\
\hline 
& \multicolumn{9}{c|}{ (e) Location and complex scale mixed change }  \\
\hline 
$\Mg$-NN & \textbf{84(56)} & \textbf{77(50)} & \textbf{74(45)} & \textbf{98(87)} & \textbf{96(84)} & \textbf{92(78)} & \textbf{78(59)} & \textbf{66(48)} & \textbf{54(39)}\\
GET & 68(45) & 60(37) & 54(31) & \textbf{93}(80) & 80(64) & 57(43) & 49(34) & 31(20) & 20(11)\\
MET & 65(38) & 58(30) & 53(27) & \textbf{94}(78) & 85(67) & 67(50) & 60(43) & 46(29) & 34(18)\\
SWR & 47(24) & 42(22) & 42(21) & 65(41) & 68(47) & 64(43) & 29(13) & 29(13) & 30(13)\\
DM & 8(0) & 8(0) & 7(0) & 3(0) & 5(0) & 5(0) & 5(0) & 4(0) & 5(0)\\
ED & 40(25) & 33(17) & 25(12) & 90(72) & 62(46) & 22(15) & 6(1) & 4(0) & 4(1)\\
$C_{2N}$ & 40(20) & 28(11) & 22(8) & 61(40) & 22(10) & 10(3) & 5(0) & 5(0) & 6(0)\\
$S_3$ & 6(0) & 8(1) & 6(0) & 4(0) & 5(0) & 5(0) & 5(0) & 4(0) & 5(0)\\
\hline 
\end{tabular}
\end{table*}

\begin{table*}
\small
    \centering
 \caption{The empirical power (detection accuracy) in percent under Settings IV-VI. Here $\Mg$-NN is the proposed method. }
    \label{tab p2} 
       \begin{tabular}{|c|ccc|ccc|ccc|}
\hline
 & \multicolumn{3}{c|}{Setting IV ($\chi_5^2$)} & \multicolumn{3}{c|}{Setting V (Gaussian Mixture) }  & \multicolumn{3}{c|}{Setting VI (Gaussian outlier)}\\ 
$d$ & 200 & 500 & 1000 & 200 & 500 & 1000 & 200 & 500 & 1000\\
  \hline 
 & \multicolumn{9}{c|}{ (a) Location change }  \\
\hline 
$\Mg$-NN & 74(56) & 65(46) & 54(37) & 32(18) & 42\textbf{(27)} & 58\textbf{(42)} & 61{(44)} & 50(34) & 42\textbf{(24)}\\
GET & 60(43) & 46(30) & 33(21) & 29(17) & 39(25) & 51(34) & 44(30) & 30(17) & 21(9)\\
MET & 64(47) & 52(35) & 39(25) & 30(18) & 41\textbf{(28)} & 54(37) & 48(33) & 34(19) & 27(13)\\
SWR & 20(7) & 18(6) & 15(3) & 20(6) & 24(9) & 31(13) & 18(7) & 15(4) & 13(4)\\
DM & 6(0) & 5(0) & 5(0) & 7(0) & 6(0) & 7(0) & 4(0) & 6(0) & 7(0)\\
ED & \textbf{94(80)} & \textbf{93(79)} & \textbf{91(76)} & 6(1) & 5(1) & 6(1) & \textbf{93}(8) & \textbf{90(73)} & \textbf{87}(7)\\
$C_{2N}$ & \textbf{95(80)} & \textbf{92(78)} & \textbf{88(73)} & \textbf{87(53)} & \textbf{83}(24) & \textbf{84}(11) & {78\textbf{(61)}} & 44(26) & 19(6)\\
$S_3$ & 5(1) & 4(0) & 6(1) & 7(0) & 6(0) & 7(0) & 4(0) & 4(0) & 6(0)\\
\hline
& \multicolumn{9}{c|}{(b) Simple scale change }  \\ 
\hline 
$\Mg$-NN & \textbf{90(64)} & {89(61)} & {86(58)} & \textbf{71(49)} & \textbf{81(58)} & \textbf{88(69)} & \textbf{84(61)} & 77(52) & {70(47)}\\
GET & 85(57) & 84(53) & 80(51) & 65(40) & 76(51) & \textbf{84}(60) & \textbf{87(61)} & \textbf{83(57)} & \textbf{75(51)}\\
MET & \textbf{88}(61) & 87(57) & 83(54) & 66(45) & 76\textbf{(55)} & 83(64) & \textbf{85(59)} & \textbf{78}(51) & \textbf{72}(44)\\
SWR & 5(1) & 6(0) & 6(0) & 5(0) & 5(0) & 5(1) & 5(0) & 5(0) & 5(1)\\
DM & \textbf{90(65)} & 82(52) & 55(25) & 6(0) & 5(0) & 5(0) & 69(50) & 59(40) & 43(25)\\
ED & 6(1) & 8(2) & 6(2) & 5(1) & 4(1) & 5(1) & 11(4) & 13(4) & 11(3)\\
$C_{2N}$ & 86(57) & \textbf{91}(63) & \textbf{91}(63) & 6(1) & 6(1) & 6(0) & 65(40) & 56(31) & 42(16)\\
$S_3$ & \textbf{93(68)} & \textbf{96(72)} & \textbf{96(72)} & 6(0) & 5(0) & 5(0) & 53(38) & 39(26) & 24(14)\\
\hline
 & \multicolumn{9}{c|}{(c) Complex scale  change }  \\
\hline 
$\Mg$-NN &  \textbf{74(49)} &  \textbf{62}(36) & 58(33) & 49(31) & 42(28) & 47(30) & \textbf{72(50)} & \textbf{69(50)} & \textbf{66(47)}\\
GET & 57(40) & 44(26) & 37(20) & 24(13) & 18(9) & 21(10) & 44(28) & 40(25) & 37(24)\\
MET & 52(29) & 41(20) & 36(15) & 21(9) & 15(6) & 17(6) & 45(27) & 43(26) & 44(25)\\
SWR & 64(43) &  \textbf{63(43)} & \textbf{64} \textbf{(42)} & \textbf{92(77)} & \textbf{92(79)} & \textbf{94(81)} & 58(36) & 57(33) & 57(34)\\
DM & 4(0) & 4(0) & 4(0) & 5(0) & 5(0) & 6(0) & 5(1) & 4(0) & 4(0)\\
ED & 8(1) & 7(1) & 8(1) & 5(0) & 5(1) & 4(1) & 8(1) & 7(1) & 7(0)\\
$C_{2N}$ & 3(0) & 4(1) & 5(0) & 5(0) & 5(0) & 6(0) & 5(0) & 5(0) & 6(0)\\
$S_3$ & 4(0) & 4(0) & 5(0) & 5(0) & 5(0) & 6(0) & 5(0) & 5(0) & 5(0)\\
\hline
& \multicolumn{9}{c|}{ (d) Location and simple scale mixed change }  \\
\hline 
$\Mg$-NN & 70(42) & 67(39) & 63(38) & \textbf{68(45)} & \textbf{81(56)} & \textbf{86(63)} & 86(62) & 78(52) & 74(47)\\
GET & 66(37) & 62(32) & 58(31) & \textbf{70(45)} & \textbf{81(56)} & \textbf{87(65)} & \textbf{93(71)} & \textbf{86(62)} & \textbf{80(60)}\\
MET & 66(39) & 66(36) & 60(34) & \textbf{67(44)} & \textbf{79(57)} & \textbf{84(62)} & \textbf{88}(64) & {81}(52) & 75(50)\\
SWR & 6(1) & 5(0) & 6(0) & 7(1) & 6(1) & 7(1) & 7(1) & 8(2) & 9(1)\\
DM & {76(46)} & 57(27) & 34(9) & 7(0) & 6(0) & 6(0) & 73(50) & 59(37) & 44(26)\\
ED & 14(5) & 10(3) & 8(3) & 5(1) & 5(1) & 4(1) & 44(26) & 38(22) & 37(20)\\
$C_{2N}$ & 73(41) & 77(44) & 79(46) & 52(22) & 38(10) & 38(5) & 76(51) & 57(32) & 43(22)\\
$S_3$ & \textbf{82(54)} & \textbf{85(55)} & \textbf{85(56)} & 6(0) & 6(0) & 6(0) & 55(38) & 36(22) & 24(14)\\
\hline 
& \multicolumn{9}{c|}{(e) Location and complex scale mixed change }  \\
\hline 
$\Mg$-NN &  \textbf{52(30)} &  \textbf{41(23)} & \textbf{39(21)} & 42(25) & 40(24) & 45(29) & \textbf{81(64)} & \textbf{76(59)} & \textbf{74(56)}\\
GET & 39(22) & 25(12) & 24(11) & 19(9) & 19(7) & 20(10) & 58(42) & 49(32) & 43(28)\\
MET & 37(16) & 26(11) & 25(10) & 15(6) & 16(5) & 18(7) & 60(41) & 52(34) & 48(30)\\
SWR & 42(22) &  \textbf{40(21)} & \textbf{40(19)} & \textbf{74(53)} & \textbf{76(54)} & \textbf{79(60)} & 56(33) & 53(32) & 50(29)\\
DM & 4(0) & 4(0) & 5(0) & 5(0) & 5(0) & 5(0) & 4(0) & 5(0) & 4(0)\\
ED & 10(2) & 7(1) & 7(1) & 4(0) & 5(1) & 6(1) & 43(25) & 34(19) & 28(14)\\
$C_{2N}$ & 8(2) & 6(1) & 5(0) & 8(1) & 9(1) & 11(1) & 20(9) & 11(2) & 4(0)\\
$S_3$ & 4(0) & 4(0) & 4(0) & 5(0) & 5(0) & 5(0) & 4(0) & 6(0) & 4(0)\\
\hline 
\end{tabular}
\end{table*}

For the multivariate $t_5$ and Cauchy distributions, $\Mg$-NN shows the highest power under the alternatives (a), (b), (d), and (e). SWR performs the best for the complex scale alternative (c), followed immediately by $\Mg$-NN, while GET and MET also have moderate power. On the contrary, DM, ED, $C_{2N}$ and $S_3$ fail for most of the alternatives under the multivariate $t_5$ and Cauchy distributions with the power close to the nominal level. It shows that $\Mg$-NN is robust to heavy-tailed distributions, while other methods such as $C_{2N}$ and $S_3$ are not. 

From Table \ref{tab p2}, we see that ED and $C_{2N}$ perform the best for the location alternative (a) under the multivariate  $\chi_5^2$ distribution, while $\Mg$-NN performs the second best and $S_3$ has no power. For the scale alternative (b), $S_3$ exhibits the highest power, and $\Mg$-NN also performs well. In addition, under the same distribution, $\Mg$-NN and SWR outperform other methods for alternatives (c) and (e). However, SWR is powerless for alternative (d) while $\Mg$-NN shows good performance. 

For the Gaussian mixture distribution, $C_{2N}$ has the highest power for the location alternative (a), while $\Mg$-NN is the second best. For alternatives (b) and (d), $\Mg$-NN has the best performance, together with GET and MET, while all other methods have unsatisfactory performance. For alternatives (c) and (e), SWR achieves the highest power, while $\Mg$-NN is also good with performance better than other methods. 

For the multivariate Gaussian distribution with $t_7$ outliers, ED is the best for the location alternative, while for $d=1000$, it is outperformed by $\Mg$-NN in the detection accuracy. For other alternatives, $\Mg$-NN almost dominates other methods, followed by GET and MET. It shows that $\Mg$-NN is robust to outliers.

In summary, the distance-based methods ED, $C_{2N}$, and $S_3$, as well as DM, are powerful for the light-tailed distribution. Specifically, ED exhibits superior power for the location alternative, $S_3$ and DM are more powerful for the simple scale alternative, while $C_{2N}$ covers both the location and the scale alternatives. Nevertheless, these methods suffer from outliers and are less powerful for heavy-tailed distributions. On the contrary, the graph-based methods GET and MET are less sensitive to outliers and show good performance for the complex scale alternative. The problem with these methods is that they use less information than distance-based methods, thus suffering from the lack of power for light-tailed distribution and the location alternative. In particular, SWR uses the least information compared to GET and MET, so it has almost no power in many settings and alternatives when other methods attain moderate power. On the other hand, $\Mg$-NN possesses good power for light-tailed distributions and shows robustness for heavy-tailed distributions.

\section{Real data examples}
\label{sec: real data}

\subsection{Seizure detection from functional connectivity networks}

\begin{figure}%[ht]
  \centering
 \includegraphics[width=3.3in]{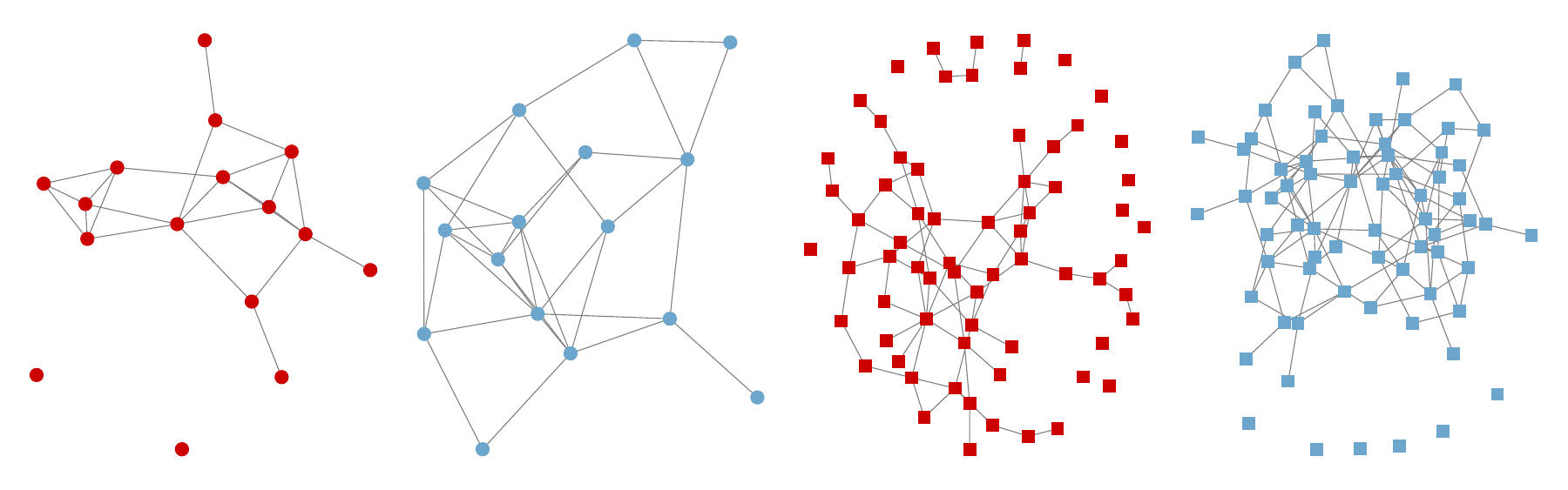}
 \caption{The functional connectivity networks of a dog (circle) and a human (square) during the period of the seizure (red) and the normal period (blue). The networks are drawn by only keeping the edges with weights larger than $0.2$. }
\label{fig:brain} 
\end{figure}

We illustrate RING-CPD for identifying epileptic seizures, which over two million Americans are suffering from \cite{iasemidis2003epileptic}. As a promising therapy, responsive neurostimulation requires automated algorithms to detect seizures as early as possible. Besides, to identify seizures, physicians have to review abundant electro-encephalogram
(EEG) recordings, which in some patients may be quite subtle. Hence, it is important to develop methods with low false positive and false negative rates to detect seizures from the EEG recordings. We use the “Detect seizures in
intracranial EEG (iEEG) recordings” database by the UPenn and Mayo
Clinic (https://www.kaggle.com/c/seizure-detection), which consists of the EEG recordings of $12$ subjects (eight patients and four dogs). For each subject, both the normal brain activity and the seizure activity are recorded multiple times, which are one-second clips with various channels (from $16$ to
$72$), reducing to a multivariate stream of iEEGs. Following the preprocessing procedure of \cite{zambon2019change}, we represent the iEEG data as functional connectivity networks using Pearson correlation in the high-gamma band ($70$-$100$Hz) \cite{bastos2016tutorial}. Functional connectivity networks are weighted graphs, where the vertices are the electrodes, and the weights of edges correspond to the coupling strength of the vertices. An illustration of the networks is in Figure~\ref{fig:brain}. 
The sample sizes of the $12$ subjects are also different, and the true change-points $\tau$ are also known - before the change-point, the networks are from the seizure period, while after the change-point, the networks are from the normal brain activity, so we have the ground truth. We use the Frobenius norm to measure the distance between the observations represented by the weighted adjacency graphs. 

We do not include SWR in the comparison here since SWR does not perform well in the simulation studies and is time-consuming. For $C_{2N}$, since it does well under some simulation settings, we try to include it in the comparison. However, it is not only time-consuming but also memory-consuming (e.g., it requires at least $17$Gb size of memory when $n=1320$); we are only able to run it for $n \leq 600$, thus only showing its result for Dog $1$, and Patients $1$ and $4$. Since the sample size of each subject is large enough, we use the asymptotic $p$-value approximation for $\Mg$-NN and MET. We omit the result of GET since it performs similarly to MET but its $p$-value approximation is not as exact as MET. For DM, ED, and $S_3$, we still use $1,000$ random permutations to obtain the $p$-values. The results are summarized in Table \ref{tab:network}, where the absolute difference between the true change-point and the detected change-point $|\hat \tau - \tau|$ is reported. The $p$-values are not reported as they are smaller than $0.01$ for all methods and subjects. Our method achieves the same detection error as MET, which is very small for all subjects. ED also performs well, but with a slightly large error for Patient $4$. Although DM and $S_3$ achieve small errors for most subjects, they attain large detection errors for Patients $3$ and $4$. The performance of $C_{2N}$ is not robust in that it shows a large detection error for Patient $4$.

\begin{table}[]
    \centering
      \caption{The absolute difference between the true change-point and the detected change-point ($|\hat \tau - \tau|$). The $p$-values of all methods for all subjects (dogs or PTs, where PT$=$ patient) are smaller than $0.01$. $\Mg$-NN is the proposed method.}
    \begin{tabular}{c|cc| cc cc cc }
    \hline 
      & $n$ & $\tau$ &  $\Mg$-NN & MET &  DM & ED & $C_{2N}$ & $S_3$  \\ \hline 
     Dog 1  & 596 & 178 &   0 & 0 &  0 & 1 & 0 & 0\\
Dog 2  & 1320 & 172 &   4 & 4 &  1 & 3 & - & 1\\
Dog 3  & 5240 & 480 &   0 & 0  &   1 & 1 &- & 1 \\
Dog 4  & 3047 & 257 &  3 & 3 &  3 & 2 & - & 3\\
\hline 
PT 1  & 174 & 70  & 1 & 1 &  1 & 0 & 7 & 1\\
PT 2  & 3141 & 151  & 7 & 7 &  13 & 1 & - & 13\\
PT 3  & 1041 & 327  & 0 & 0 &  162 & 1 & - & 162\\
PT 4  & 210 & 20  & 0 & 0 &  67 & 11 & 137 & 67\\
PT 5 & 2745 & 135  & 3 & 3 &  5 & 2 & - & 5 \\
PT 6  & 2997 & 225  & 0 & 0 &0  & 1 & - & 0\\
PT 7  & 3521 & 282  & 2 & 2 & 3 & 4 & - & 3\\
PT 8  & 1890 & 180  & 0 &  0 & 0 & 1 & - & 0\\
\hline
    \end{tabular}
    \label{tab:network}
\end{table}

\subsection{Change-point detection for MNIST handwritten digits}
\label{mnist}

\begin{table*}[]
\centering
 \caption{The empirical power with the   average change-point estimation error 
 in parentheses. $\Mg$-NN is the proposed method. The method with the highest empirical power is highlighted in bold. If two methods have the same power, choose the one with the smaller average change-point estimation error.}
\begin{tabular}{ccccccccc}
  \hline
& Setting & $\Mg$-NN & GET & MET & SWR & DM  & $C_{2N}$ & $S_3$ \\ 
  \hline
\multirow{5}{*}{$(\tau,n) = (15,30)$} & 0/8 & \textbf{1.00 (0.06)}  & 1.00 (0.10)   & 1.00 (0.10)   & 0.99 (0.09) & 0.90 (0.20) & 1.00 (4.38) & 0.96 (0.17)   \\ 
&  1/7 & \textbf{1.00 (0.07)} & \textbf{1.00 (0.07)}  & \textbf{1.00 (0.07)} & 1.00 (0.10) & 0.99 (0.13) & 1.00 (3.29) & 1.00 (0.10) \\ 
 &5/6 & \textbf{1.00 (0.20)}  & 1.00 (0.43) & 1.00 (0.43) & 0.99 (0.29)  & 0.37 (2.79) & 1.00 (2.79) & 0.30 (1.60) \\ 
 & 3/8 & \textbf{0.98 (0.89)} & 0.93 (0.98) & 0.96 (1.04) & 0.78 (1.17)  & 0.22 (8.91) & 0.96 (2.16) & 0.15 (4.00) \\ 
 & 4/9 & \textbf{0.61 (1.08)} & 0.54 (1.78) & 0.51 (1.43) & 0.56 (2.34) & 0.13 (10.62) &  0.62 (2.39) & 0.04 (13.00) \\ 
   \hline
   \multirow{5}{*}{$(\tau,n) = (15,45)$} & 0/8 & \textbf{1.00 (0.02)} & 1.00 (0.05)  & 1.00 (0.05) & 1.00 (0.16)  & 1.00 (0.38)   & 1.00 (8.90)  & 1.00 (0.20) \\ 
&  1/7 &  1.00 (0.09) & 1.00 (0.16) & 1.00 (0.13) & \textbf{1.00 (0.07)}  &  1.00 (0.20)   & 1.00 (2.69) &  1.00 (0.17) \\ 
 &5/6 & \textbf{1.00 (0.17)}  & 1.00 (0.38) & 1.00 (0.46) & 1.00 (0.39) & 0.76 (3.72)  & 1.00 (7.02)  & 0.96 (0.75)\\ 
 & 3/8 & \textbf{1.00 (0.45)}  & 1.00 (0.55) & 1.00 (0.55)  & 0.98 (1.63)  & 0.52 (7.37) &  1.00 (4.30) & 0.81 (1.30) \\ 
 & 4/9 & \textbf{0.86 (1.55)} & 0.80 (2.33) & 0.77 (2.68) & 0.70 (2.76)  & 0.11 (16.00) &  0.79 (2.88)  &  0.18 (6.89) \\ 
   \hline
\end{tabular}
\label{res:digit}
\end{table*}

To better demonstrate the efficacy of RING-CPD on non-Euclidean datasets, we evaluated its performance using the MNIST dataset of handwritten digits \cite{lecun1998gradient}. The MNIST dataset comprises images of digits ranging from $0$ to $9$, with each image being a $28 \times 28$ matrix. Each matrix element represents a grayscale intensity value between $0$ and $255$. For our analysis, we select image pairs that are typically challenging to differentiate: digits $0$ and $8$, $1$ and $7$, $5$ and $6$, $3$ and $8$, and $4$ and $9$. We structure the dataset such that the first $\tau$ images correspond to one digit and the remaining $n - \tau$ images to another, with configurations $(\tau,n) = (15,30)$ and $(15,45)$. A nominal level $\alpha =0.05$ is used, and each scenario is repeated $100$ times through random sampling of images. The negative Frobenius norm serves as the similarity measure.

We present the empirical power (defined as the proportion of trials with a $p$-value less than $0.05$) in Table~\ref{res:digit} with the average change-point estimation error $|\tau - \widehat{\tau}|$ for those rejecting $H_0$ in parentheses. Notably, the traditional Euclidean distance-based method ED is ineffective across all scenarios, evidenced by a zero empirical detection rate, so we omit its results in Table~\ref{res:digit}. In contrast, $\Mg$-NN demonstrates superior performance in nearly all configurations. Other graph-based methodologies like GET, MET, and SWR also perform commendably. However, methods such as DM, $C_{2N}$, and $S_3$ perform unsatisfactorily in some scenarios, likely due to the sparse and complex nature of image data.

\subsection{Changed interval detection for New York taxi data}
\begin{figure*}[!t]
  \centering
      \begin{tabular}{cc}
        \includegraphics[width=0.34\textwidth]{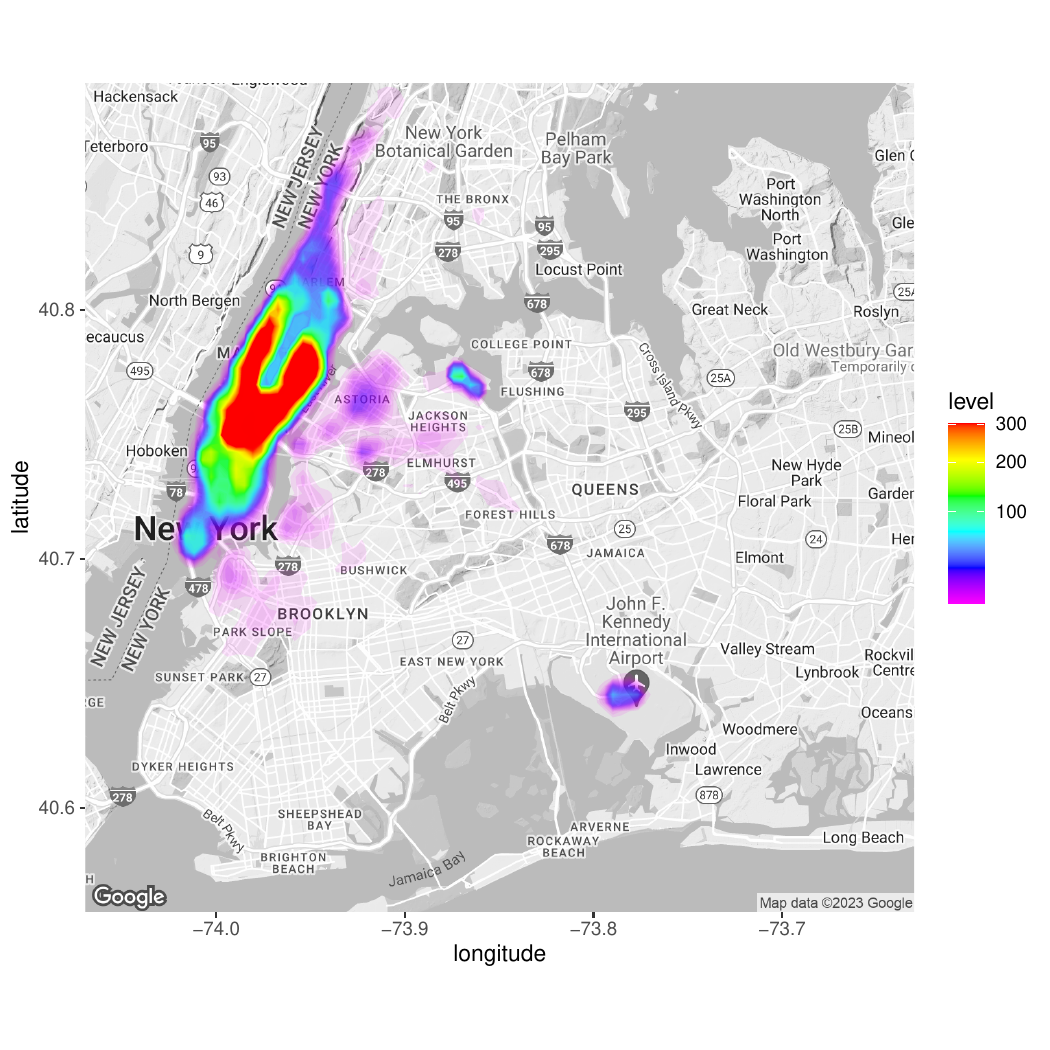}   &  \includegraphics[width=0.34\textwidth]{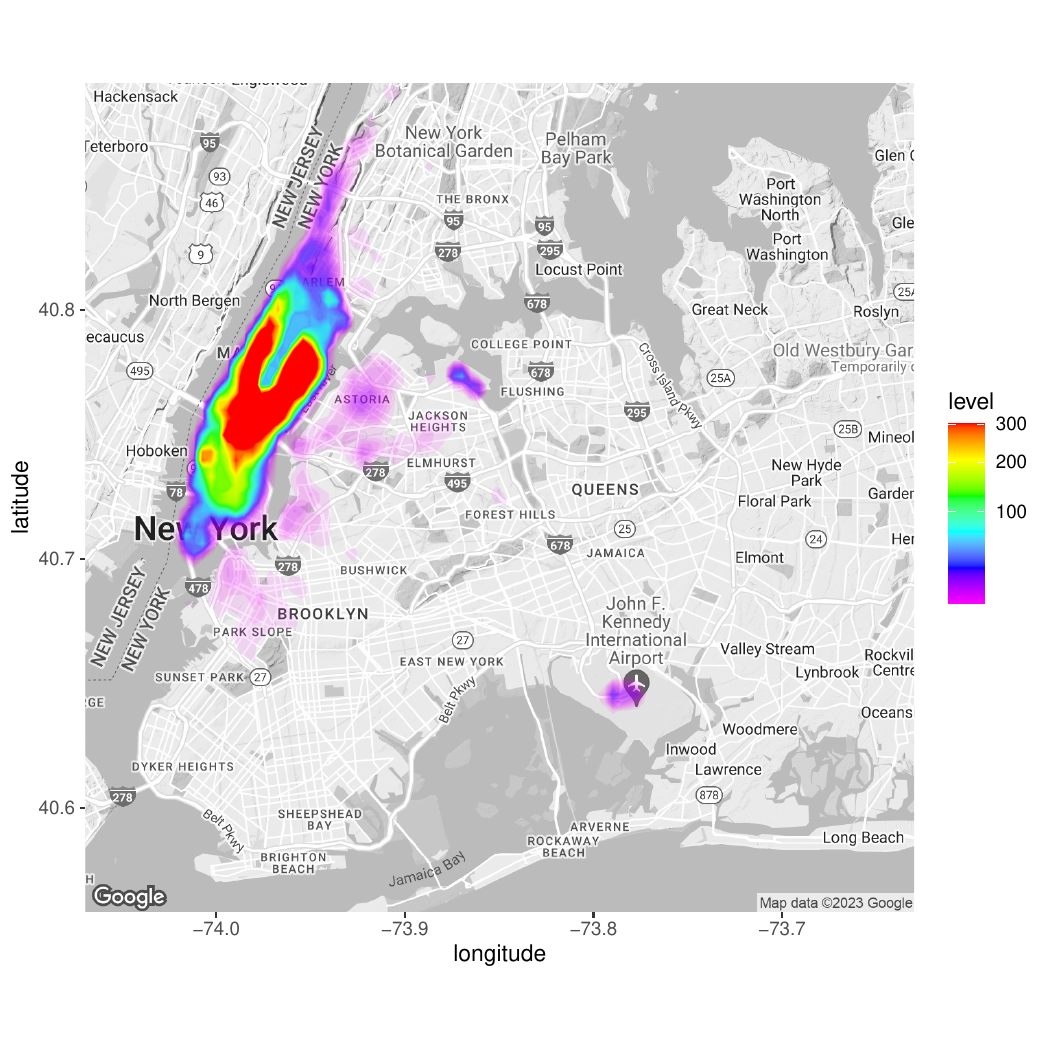}
      \end{tabular}
      \caption{Density heatmap of taxi pick-ups for dates 01/01 and 02/01 in year 2014.}
\label{fig:map}
\end{figure*}

We illustrate our method for changed interval detection in studying travel pattern changes around New York Central Park. We use the yellow taxi trip records for the year $2014$ available on the  the NYC Taxi and Limousine Commission (TLC) website (https://www1.nyc.gov/site/tlc/about/tlc-trip-record-data.page), which contains the city's taxi pickup and drop-off times and locations (longitude and latitude coordinates). We set the latitude range of New York Central Park as $40.76$ to $40.81$ and the longitude range as $-74.10$ to $-73.60$. The boundary of New York City is set as $40.50$ to $40.95$ in latitude and  $-74.10$ to $-73.60$. We only consider those trips that began with a pickup in  New York City and ended with a drop-off in New York Central Park. We use the two-dimensional kernel density estimation with the bivariate normal kernel and $50$ grid points in each direction to represent the trips of each day in New York City, as illustrated by Figure~\ref{fig:map} on two random days. We use the Frobenius norm to construct the similarity graphs in the subsequent analysis. The $p$-values of all methods are obtained through $1000$ random permutations. 

\begin{table*}[!b]
\small
    \centering
      \caption{The detected changed intervals and corresponding $p$-values of $\Mg$-NN, GET and MET for the NYC taxi data. }
    \begin{tabular}{c|cccc}
    \hline 
     Time period &  $\Mg$-NN & GET & MET & Nearby Events \\ \hline $01/01$-$12/31$ &  \multicolumn{3}{c}{ \textbf{06/19-08/31}} & \multirow{2}{*}{Summer break}\\
$p$-value & \multicolumn{3}{c}{$<0.001$} \\
  \hline 
$01/01$-$06/18$ &   \multicolumn{3}{c}{ \textbf{03/18-03/28}} &  \multirow{2}{*}{Spring break/St. Patrick's Day} \\ 
$p$-value & \multicolumn{3}{c}{$<0.001$} \\
$06/19$-$08/31$ & $06/23$-$06/27$ & $07/17$-$08/29$ & $06/23$-$06/27$ & \multirow{2}{*}{-} \\
$p$-value & $0.048$ & $0.196$ & $0.223$ \\
$09/01$-$12/31$ &  \textbf{09/01-11/12} 
& \textbf{09/03-11/12} & \textbf{09/02-11/12} & \multirow{2}{*}{Fall Semester (till Veterans Day)} \\
$p$-value & \multicolumn{3}{c}{$<0.001$} \\
 \hline 
$01/01$-$03/17$  & \textbf{01/27-01/31} & $02/19$-$03/13$ & $02/19$-$03/13$ & \multirow{2}{*}{Lunar New Year} \\
$p$-value & $0.01$ & $0.051$ & $0.263$  \\
$03/29$-$06/18$ & $04/04$-$06/15$ & $04/23$-$05/22$  & $04/23$-$05/23$  & \multirow{2}{*}{-} \\
$p$-value & $0.094$ & $0.366$ & $0.333$ \\
$09/01(03,02)$-$11/12$  & $09/02$-$09/11$  & $09/03$-$09/11$ & $09/02$-$09/12$  &  \multirow{2}{*}{-}\\ 
$p$-value & $0.014$ & $0.073$ & $0.028$ \\
$11/13$-$12/31$ & $12/22$-$12/30$ & $12/20$-$12/30$ & $12/20$-$12/30$ &  \multirow{2}{*}{-}\\
$p$-value & $0.020$ & $0.067$ & $0.059$\\
  \hline 
 $02/01$-$03/17$ &  $03/03$-$03/07$ 
 & & & 
 \multirow{2}{*}{-}\\
$p$-value & $0.263$ & 
\\ 
 \hline 
    \end{tabular}
    \label{tab:taxi0}
\end{table*}

We first compare RING-CPD with GET and MET. We set $n_0 = \max\{5,[0.05n]\}$, $n_1 = n - n_0$, and 
the nominal level to be $0.01$. All methods detect the same changed interval $06/19$-$08/31$ with $p$-values $<0.001$, which almost entirely overlaps with the summer break. Besides, $06/19$ is Juneteenth and $09/01$ is Labor Day.

Since there may be multiple changed intervals, we apply the methods recursively. Specifically, we apply the methods to the three segments divided by the detected changed interval. For the period of $01/01$-$06/18$, all methods detect the changed interval $03/18$-$03/28$ with $p$-value $<0.001$, which is around the spring break period for most American universities, while $3/17$ is St. Patrick's Day. For the period of $06/19$-$08/31$, no method can reject the null hypothesis at the $0.01$ significance level. For the period of $09/01$-$12/31$, $\Mg$-NN, GET, and MET report the changed interval starting at $09/01$, $09/03$, and $09/02$ respectively and all ending at $11/12$ with $p$-values $<0.001$. This is the changed interval of the Fall Semester till Veterans Day ($11/11$). 

We further apply these methods to the segments  that are longer than $40$ days. The only detected changed interval is $01/27$-$01/31$ reported by $\Mg$-NN with $p$-value $0.01$ in the segment $01/01$-$03/17$, which is around the Lunar New Year ($01/31/2014$). It is worth noting that for the period of $11/13$-$12/31$, all methods report the changed interval that covers Christmas, while $\Mg$-NN yields a small $p$-value $0.02$ and other methods report  $p$-values larger than $0.05$. The results are summarized in Table \ref{tab:taxi0}. 

We also apply other methods to the dataset. Since both ED and  $S_3$ can detect multiple change points, we apply them directly to the whole sequence $01/01$-$12/31$. We also include $C_{2N}$. Although $C_{2N}$ is not designed for multiple change-point detections, we adopt the same binary segmentation procedure used by $S_3$ \cite{nie2021weighted}.   
As summarized in Table~\ref{tab:taxi1}, ED detects two change-points $06/20$ and $09/02$ both with $p$-values $0.001$. $S_3$ only detects one change-point $06/20$ with $p$-value $0.003$.
$C_{2N}$ detects two change-points, which are $06/19$ and $09/14$, with $p$-values $<0.001$ and $0.004$. They all miss some important change points shown in Table \ref{tab:taxi1}.

\begin{table}[!t]
\small
    \centering
      \caption{The detected change-points and corresponding $p$-values of ED, $S_3$ and $C_{2N}$ for the NYC taxi data. }
    \begin{tabular}{ c|cc |c | cc}
    \hline 
   \multicolumn{1}{c}{Method}  & \multicolumn{2}{c}{ED}  & \multicolumn{1}{c}{$S_3$}  & \multicolumn{2}{c}{$C_{2N}$} \\
   \hline 
    CP  & $06/20$ & $09/02$  &  $06/20$    & $06/19$ & $09/14$ \\
 $p$-value  &  $0.001$ & $0.001$ &  $0.003$  & $<0.001$ & $0.004$ \\
\hline 
    \end{tabular}
    \label{tab:taxi1}
\end{table}

\section{Discussion}

\subsection{Kernel and Distance IN Graph CPD}

The proposed approach can be extended to use weights other than ranks in weighting the edges in the similarity graph. For example, we could incorporate kernel value or distance directly to have Kernel IN  Graph Change-Point Detection (KING-CPD) and Distance IN Graph Change-Point Detection (DING-CPD) methods. Specifically, we can use the kernel values or (negative) pairwise distances to weight the similarity graph, and the asymptotic property still holds under mild conditions. Let 
\[ K_{i j } = K(y_i,y_j) \indi\big( (i,j) \in G_k \big) \,,\]
where $K$ is a kernel function or a negative distance function, for example, the Gaussian kernel $K(y_i,y_j) = \exp \big(-\|y_i - y_j\|^2/(2 \sigma^2) \big)$ with the kernel bandwidth $\sigma$ or $K(y_i,y_j)$ simply the negative $l_1$ distance $K(y_i,y_j) = -\|y_i - y_j\|_1$, and 
$G_k$ is a similarity graph such as the $k$-NNG or the $k$-MST. We require Condition (7) $\max_{i,j} K_{ij} = o\big( n^2 r_d^2 \big)$, which essentially states that there is no element of $K_{ij}$ dominates others.

\begin{theorem}
Replacing $R_{i j}$ by $K_{i j}$ in Conditions (1)-(6), and the definition of $Z_{\diff}$ and $Z_w$, then under Conditions (1)-(6) and (7), we have
\begin{enumerate}
    \item $\big\{Z_{\diff}( \lfloor n u \rfloor): 0<u<1\big\}$ and $\big\{Z_{w}( \lfloor n u \rfloor): 0<u<1\big\}$ converge to independent Gaussian processes in finite-dimensional distributions, which we denote as $\big\{Z_{\diff}^{*}(u): 0<u<1\big\}$ and $\big\{Z_{w}^{*}(u): 0<u<1\big\}$, respectively.
    \item  $\big\{Z_{\diff}(\lfloor n u \rfloor, \lfloor n v \rfloor): 0<u<v<1\big\}$ and $\big\{Z_{w}(\lfloor n u \rfloor, \lfloor n v \rfloor): 0<u<v<1\big\}$ converge to independent two-dimension Gaussian random fields in finite-dimensional distributions, which we denote as $\big\{Z_{\diff}^{*}(u, v): 0<u<v<1\big\}$ and $\big\{Z_{w}^{*}(u, v): 0<u<v<1\big\}$, respectively.
\end{enumerate}
\label{thm:king}
In addition, Theorem \ref{thm:cov} also holds by replacing $R_{ij}$ by $K_{ij}$. 
\end{theorem}

The proof of Theorem \ref{thm:king} follows straightforwardly from the proof of Theorems \ref{thm:basis} and \ref{thm:cov}, thus omitted here.

\section*{Acknowledgments}
The authors would like to thank the associate editor and referees for their constructive comments and suggestions.

\bibliographystyle{IEEEtran}
\bibliography{ref}

\begin{IEEEbiographynophoto}{Doudou Zhou}
 received his Ph.D. in Statistics from the University of California, Davis, in 2022. From 2022 to 2024, he was a Postdoctoral Researcher at Harvard University. He is currently an Assistant Professor in the Department of Statistics and Data Science at the National University of Singapore. His research focuses on developing theoretical and computational methods for high-dimensional data analysis and machine learning.   
\end{IEEEbiographynophoto}

\begin{IEEEbiographynophoto}{Hao Chen}
received her Ph.D. in Statistics from Stanford University in 2013. Since then, she joined the Department of
Statistics, University of California, Davis, as a faculty member, where she is currently an Associate Professor (with tenure). Her research interests include developing practical and robust methods that can
deal with various data types, including high-dimensional data, image data,
and network data, for hypothesis testing, signal detection, classification, and
clustering.
\end{IEEEbiographynophoto}
% \newpage

% \section{Biography Section}
% If you have an EPS/PDF photo (graphicx package needed), extra braces are
%  needed around the contents of the optional argument to biography to prevent
%  the LaTeX parser from getting confused when it sees the complicated
%  $\backslash${\tt{includegraphics}} command within an optional argument. (You can create
%  your own custom macro containing the $\backslash${\tt{includegraphics}} command to make things
%  simpler here.)
 
% \vspace{11pt}

% \bf{If you include a photo:}\vspace{-33pt}
% \begin{IEEEbiography}[{\includegraphics[width=1in,height=1.25in,clip,keepaspectratio]{fig1}}]{Michael Shell}
% Use $\backslash${\tt{begin\{IEEEbiography\}}} and then for the 1st argument use $\backslash${\tt{includegraphics}} to declare and link the author photo.
% Use the author name as the 3rd argument followed by the biography text.
% \end{IEEEbiography}

% \vspace{11pt}

% \bf{If you will not include a photo:}\vspace{-33pt}
% \begin{IEEEbiographynophoto}{John Doe}
% Use $\backslash${\tt{begin\{IEEEbiographynophoto\}}} and the author name as the argument followed by the biography text.
% \end{IEEEbiographynophoto}

\vfill

\end{document}